\documentclass[12pt]{article}

\usepackage{graphicx}

\usepackage{lmodern}
\usepackage[T1]{fontenc}
\usepackage{textcomp}

\usepackage[intlimits]{amsmath}
\usepackage{amssymb}
\usepackage{bbm}
\usepackage{mathrsfs}
\usepackage{axodraw}
\usepackage{wrapfig}
\usepackage{caption}
\usepackage{subfigure}
\usepackage{enumerate}
\usepackage{epsfig}
\usepackage{cite}

\def\bea{\begin{eqnarray}}
\def\eea{\end{eqnarray}}

\newcommand{\jvs}{\rule[-7pt]{0.00pt}{20pt}}

\def\e{\epsilon}

\def\m{\mu}
\def\n{\nu}

\def\z{\zeta}

\def\F{\Phi}

\def\ve{\varepsilon}

\def\T{{\bf 10}}
\def\F{{\bf 5}}

\def\1{{\bf 1}}

\def\Mp{{M_{\rm P}}}
\def\mG{{m_{3/2}}}

\def\bo{{\raise.15ex\hbox{\large$\Box$}}}               
\def\face{{\raise.2ex\hbox{$\displaystyle \bigodot$}\mskip-2.2mu \llap {$\ddot
        \smile$}}}                                      
\def\dg{\dagger}                                     

\def\leftrightarrowfill{$\mathsurround=0pt \mathord\leftarrow \mkern-6mu
        \cleaders\hbox{$\mkern-2mu \mathord- \mkern-2mu$}\hfill
        \mkern-6mu \mathord\rightarrow$}       
\def\dvec#1{\vbox{\ialign{##\crcr
        \leftrightarrowfill\crcr\noalign{\kern-1pt\nointerlineskip}
        $\hfil\displaystyle{#1}\hfil$\crcr}}}           

\def\beq{\begin{equation}}
\def\eeq{\end{equation}}

\def\lsim{\mathrel{\raise.3ex\hbox{$<$\kern-.75em\lower1ex\hbox{$\sim$}}}}
\def\gsim{\mathrel{\raise.3ex\hbox{$>$\kern-.75em\lower1ex\hbox{$\sim$}}}}

\begin{document}
\date{\mbox{ }}

\title{
{\normalsize
DESY 10-068\hfill\mbox{}\\
July 2010\hfill\mbox{}\\}
\vspace{1cm}
\bf Broken R-Parity\\ in the Sky and at the LHC \\[8mm]}
%
\author{S.~Bobrovskyi, W.~Buchm\"uller, J.~Hajer, J.~Schmidt \\[2mm]
{\normalsize\it Deutsches Elektronen-Synchrotron DESY, Hamburg, Germany}}
\maketitle

\thispagestyle{empty}

\begin{abstract}
\noindent
Supersymmetric extensions of the Standard Model with small R-parity
and lepton number violating couplings are naturally consistent with
primordial nucleosynthesis, thermal leptogenesis and gravitino dark matter.
We consider supergravity models with universal boundary conditions at the
grand unification scale, and scalar $\tau$-lepton or bino-like neutralino as
next-to-lightest superparticle (NLSP). Recent Fermi-LAT data on the isotropic
diffuse gamma-ray flux yield a lower bound on the gravitino lifetime.
Comparing two-body gravitino and neutralino decays we find a lower bound
on a neutralino NLSP decay length, $c\tau_{\chi^0_1} \gsim 30~\mathrm{cm}$.
Together with gravitino and neutralino masses one obtains
a microscopic determination of the Planck mass.
For a $\widetilde\tau$-NLSP there exists no model-independent
lower bound on the decay length. Here the strongest bound comes from the
requirement that the cosmological baryon asymmetry is not washed out, which
yields $c\tau_{\widetilde\tau_1} \gsim 4~\mathrm{mm}$. However, without
fine-tuning of parameters, one finds much larger decay lengths. For
typical masses, $m_{3/2} \sim 100~\mathrm{GeV}$ and
$m_{\mathrm{NLSP}} \sim 150~\mathrm{GeV}$, the discovery of a photon
line with an intensity close to the Fermi-LAT limit would imply
a decay length $c\tau_{\mathrm{NLSP}}$ of several hundred meters, which
can be measured at the LHC.

\end{abstract}

\newpage

\section{Introduction}

Locally supersymmetric extensions of the Standard Model predict the existence
of the gravitino, the gauge fermion of supergravity \cite{fnf76}. For some
patterns of supersymmetry breaking, the gravitino is the lightest
superparticle (LSP), and therefore a natural dark matter candidate \cite{pp81}.
Heavy unstable gravitinos may cause the `gravitino problem'
\cite{we82,ens85,kkm05} for large reheating temperatures in the early universe.
This is the case for thermal leptogenesis \cite{fy86}, where gravitino
dark matter has become an attractive alternative \cite{bbp98} to the standard
WIMP scenario \cite{fen05}.

Recently, it has been shown that models with small R-parity and lepton
number breaking naturally yield a consistent cosmology incorporating
primordial
nucleosynthesis, leptogenesis and gravitino dark matter \cite{bcx07}. The
gravitino is no longer stable, but its decays into Standard Model (SM)
particles are doubly suppressed by the Planck mass and the small R-parity
breaking parameter. Hence, its lifetime can exceed the age of the Universe
by many orders of magnitude, and the gravitino remains a viable dark matter
candidate \cite{ty00}.

Gravitino decays lead to characteristic signatures in high-energy cosmic
rays, in particular to a diffuse gamma-ray flux
\cite{bcx07,ty00,lor07,bbx07,it07,imm08,bix09,blx09,crx10}. The recent search
of the Fermi-LAT collaboration for monochromatic photon lines
\cite{FermiLAT1} and the measurement of the diffuse gamma-ray flux
up to photon energies of $100~\mathrm{GeV}$ \cite{FermiLAT2} severely
constrain possible signals from decaying dark matter. In this paper
we study the implications of this data for the decays of the
next-to-lightest superparticle (NLSP)
at the LHC, extending the estimates in \cite{bcx07}.

We shall restrict our analysis to the simplest class of supergravity models
with universal boundary conditions at the Grand Unification (GUT) scale,
which lead to neutralino or $\widetilde\tau$-NLSP. Electroweak precision
tests, thermal leptogenesis and gravitino dark matter together allow
gravitino and NLSP masses in the range $m_{3/2} = 10\ldots 500~\mathrm{GeV}$
and $m_{\rm NLSP} = 100\ldots 500~\mathrm{GeV}$ \cite{bes08}.
Following \cite{bcx07}, the breaking of R-parity is tied to the breaking
of lepton number, which leads to a model with bilinear R-parity breaking
\cite{hs84,add04}. The soft supersymmetry breaking terms are characteristic
for gravity or gaugino mediation.

In order to establish the connection between the gamma-ray flux from
gravitino decays and NLSP decays, one needs R-parity breaking matrix
elements of neutral, charged and supercurrents. For the considered
supergravity models we are able to obtain these matrix elements
to good approximation analytically. This makes our results for the
NLSP decay lengths rather transparent. As we shall see, the lower
bound on the neutralino decay length is a direct consequence of the
Fermi-LAT constraints on decaying dark matter. On the other hand, the
lower bound on the $\widetilde\tau$-decay length is determined by the
cosmological bounds on R-parity breaking couplings, which follow from
the requirement that the baryon asymmetry is not washed out
\cite{cdx91,ehi09}.

This paper is organized as follows. In Section~2 we discuss the general
Lagrangian for R-parity breaking in a basis of scalar $SU(2)$ doublets
where all bilinear mixing terms vanish. This leads to new Yukawa and
gaugino couplings, some of which are proportional to the up-quark Yukawa
couplings. Section~3 deals with the various supersymmetry, R-parity
and lepton number breaking terms in the Lagrangian and the relations
among them due to a $U(1)$ flavour symmetry of the considered model.
The needed R-parity breaking matrix elements of neutral, charged and
supercurrent are analytically calculated in Section~4, based on the
diagonalization of the mass matrices which is discussed in detail in the
appendix. The main results of the paper, the bounds on the NLSP decay
lengths and the partial decay widths, are described in Section~5,
followed by our conclusions in Section~6.\\

\section{Bilinear R-Parity Breaking}

Supersymmetric extensions of the Standard Model with bilinear R-parity
breaking contain mass mixing terms between lepton and Higgs fields in the
superpotential\footnote{Our notation
for Higgs and matter superfields, scalars and left-handed fermions reads:
$H_u = (H_u, h_u)$, $l_i = (\tilde{l}_i,l_i)$ etc.},
\begin{equation}\label{rpvw1}
   \Delta W = \mu_i H_u l_i \,
\end{equation}
as well as the scalar potential induced by supersymmetry breaking,
\begin{equation}\label{rpvv1}
 -\Delta {\cal L} = B_i H_u \tilde{l}_i
+ m^2_{id} {\tilde l}^\dagger_i H_d + \mathrm{h.c.} \, .
\end{equation}
These mixing terms,
together with the R-parity conserving superpotential
\begin{equation}\label{w1rp}
W = \mu H_u H_d + h_{ij}^{u} q_i u^c_j H_u + h_{ij}^{d} d^c_i q_j H_d
+ h_{ij}^{e} l_i e^c_j H_d \ ,
\end{equation}
the scalar mass terms
\begin{align}\label{scalarmass}
-{\cal L}_{\mathrm{M}} =
&m^2_{u} H_u^\dagger H_u + m^2_{d} H_d^\dagger H_d
+ (B H_u H_d + \mathrm{h.c.}) \nonumber \\
&+ {\widetilde m}^{2}_{li} {\tilde l}_i^\dagger {\tilde l}_i
+ {\widetilde m}^{2}_{ei} {\tilde e}_i^{c\dagger} {\tilde e}^c_i
+ {\widetilde m}^2_{qi} {\tilde q}_i^{\dagger} {\tilde q}_i
+ {\widetilde m}^2_{ui} {\tilde u}_i^{c\dagger} {\tilde u^c}_i
+ {\widetilde m}^2_{di} {\tilde d}_i^{c\dagger} {\tilde d^c}_i \ ,
\end{align}
and the standard $SU(3)\times SU(2)\times U(1)_Y$ gauge interactions define
the supersymmetric standard model with bilinear R-parity breaking. Note that
the Higgs mass terms $m_u^2$ and $m_d^2$ contain the contributions from the
superpotential (\ref{w1rp}) and the soft supersymmetry breaking terms.
For simplicity, we have assumed flavour diagonal mass matrices in
(\ref{scalarmass}).

For a generic choice of parameters the electroweak symmetry is broken by
vacuum expectation values (VEVs) of all scalar $SU(2)$ doublets,
\begin{equation}\label{vevs}
\langle H_u^0 \rangle = v_u \ , \quad
\langle H_d^0 \rangle = v_d \ , \quad
\langle \tilde{\nu}_i \rangle = v_i \ ,
\end{equation}
with\footnote{Note that our result for $\widehat\epsilon_i = v_i/v_d$
holds at all renormalization scales, contrary to different expressions
used in the literature.}
\begin{equation}\label{sneuvevs}
\frac{v_u}{v_d} \equiv \tan{\beta}\ , \quad
\widehat \epsilon_i \equiv \frac{v_i}{v_d}
  = \frac{ B_i \tan \beta - m_{id}^{2} - \mu \mu_i^* }
         {\widetilde m_{li}^2 + \frac{1}{2} m_Z^2 \cos 2 \beta } \ ,
\end{equation}
where higher order terms in the R-parity breaking parameters have been
neglected.

It is convenient to discuss the predictions of the model in a basis of
$SU(2)$ doublets where the mass mixings $\mu_i$, $B_i$ and $m_{id}^2$ in
Eqs.~(\ref{rpvw1}) and (\ref{rpvv1}) are traded for R-parity breaking
Yukawa coulings. This can easily be achieved by field redefinitions.
First one rotates the superfields $H_d$ and $l_i$,
\begin{equation}\label{rot1}
H_d = H_d' - \e_i l'_i\ , \quad l_i = l'_i + \e_i H_d'\ ,
\quad \e_i = \frac{\mu_i}{\mu} \ .
\end{equation}
Then the bilinear term (\ref{rpvw1})
vanishes for the new fields, i.e., $\mu_i' = 0$, and one obtains instead the
cubic R-parity violating terms
\begin{equation}\label{wrot}
\Delta W' = \frac{1}{2}\lambda_{ijk} l'_i e^c_j l'_k + \lambda'_{ijk} d^c_i q_j l'_k \ ,
\end{equation}
where
\begin{equation}\label{lambda}
\lambda_{ijk} =  - h_{ij}^{e}\e_k + h_{kj}^{e}\e_i \ , \quad
\lambda'_{ijk}=  - h_{ij}^{d}\e_k \ .
\end{equation}
The new R-parity breaking mass mixings are given by
\begin{equation}
B'_i = B_i-B\e_i \ , \quad
m^{2\prime}_{id} = m^2_{id} + \e_i (\widetilde m^2_{li} - m^2_{d}) \ .
\label{softnewbasis}
\end{equation}
The corrections for R-parity conserving mass terms are negligable.

In a second step one can perform a non-supersymmetric rotation among all
scalar $SU(2)$ doublets,
\begin{equation}\label{rot2}
H'_d = H''_d - \e'_i \tilde{l}^{\prime \prime}_i\ , \quad
\ve H^*_u = \ve H^{\prime *}_u - \e''_i \tilde{l}^{''}_i\ , \quad
\tilde{l}'_i = \tilde{l}''_i + \e'_i H''_d + \e''_i \ve H^{\prime *}_u \ ,
\end{equation}
where $\ve$ is the usual $SU(2)$ matrix, $\ve = i\tau^2$.
Choosing
\begin{align}
\e'_i &= -\frac{B_i^{\prime} B + m^{2\prime}_{id}\left(\widetilde{m}^2_{li}
- m_u^2\right)}
{\left(\widetilde{m}^2_{li} - m_u^2\right)
\left(\widetilde{m}^2_{li} - m_d^2\right) - B^2}\ , \label{rpv1newbasis}\\
\e''_i & =\frac{B_i'\left(\widetilde{m}^2_{li}-m_d^2\right)+B m^{2\prime}_{id}}
{\left(\widetilde{m}^2_{li} - m_u^2\right)
\left(\widetilde{m}^2_{li} - m_d^2\right) - B^2}\ , \label{rpv2newbasis}
\end{align}
the $H_u\tilde{l}_i$ and $\tilde{l}^\dagger H_d$ mixing terms vanish in the new
basis of doublets. According to (\ref{sneuvevs}) also the scalar lepton VEVs
$\langle \tilde{\nu}_i \rangle$ vanish in this basis.

It is straightforward to work out the R-parity violating Yukawa couplings
which are induced by the rotation (\ref{rot2}). We are particularly
interested in the terms containing one light superparticle, i.e, a scalar
lepton, bino or wino. The corresponding couplings read, after dropping
prime and double-prime superscripts on all fields\footnote{Our notation for
gauge fields, field strengths and left-handed gauginos reads: $B_{\mu}$,
$B_{\mu\nu}$, $b$ etc.},
\begin{align}\label{yukrpv}
-\Delta {\cal L} \supset \
&\frac{1}{2}\lambda_{ijk} l_i \tilde{e}^c_j l_k
+ \lambda'_{ijk} d^c_i q_j \tilde{l}_k
+ \hat{\lambda}_{ijk} l_i e^c_j \tilde{l}_k
+ \hat{\lambda}'_{ijk} q_i u^c_j \ve\tilde{l}^*_k \nonumber\\
&+ h^e_{ij}(\e_i'H_d + \e_i'' \ve H^*_u)e^c_j h_d \nonumber\\
& - \frac{g'}{\sqrt{2}}(\e_i'H_d^\dagger - \e_i'' H^T_u \ve)l_i b
 + \frac{g}{\sqrt{2}}(\e_i'H_d^\dagger - \e_i'' H^T_u \ve)\tau^I l_i w^I
+ \mathrm{h.c.}\ ,
\end{align}
where the Yukawa couplings are given by
\begin{align}
\lambda_{ijk} &=  - h_{ij}^{e}\e_k + h_{kj}^{e}\e_i \ , \quad
\lambda'_{ijk} =  - h_{ij}^{d}(\e_k+\e_k') \ , \label{lambda2}\\
\hat{\lambda}_{ijk} &=  - h_{ij}^{e}(\e_k+\e_k') + h_{kj}^{e}\e_i \ , \quad
\hat{\lambda}'_{ijk}=  h_{ij}^{u}\e_k'' \ . \quad \label{hatlambda}
\end{align}
Since the field transformations are non-supersymmetric, the couplings
$\lambda_{ijk}$ and $\hat{\lambda}_{ijk}$ are no longer equal as in Eq.~(\ref{lambda}).
Furthermore, a new coupling of right-handed up-quarks, $\hat{\lambda}'_{ijk}$,
has been generated.

After electroweak symmetry breaking one obtains new mass mixings between
higgsinos, gauginos and leptons,
\begin{align}
-\Delta {\cal L}_M \supset \
m^e_{ij}\frac{\z_i}{c_\beta} e^c_j h_d
 - m_Z s_w \z_i^* \nu_i b
 + m_Z c_w \z_i^* \nu_i w^3 + \mathrm{h.c.} \ ,
\end{align}
where we have defined
\begin{align}
\z_i &= \frac{\e_i'v_d + \e_i'' v_u}{v}\ , \quad
v = \sqrt{v_u^2 + v_d^2}\ , \quad \frac{v_u}{v_d} =
\tan{\beta} \equiv \frac{s_\beta}{c_\beta} \ ,
\label{zetabeta} \\
m^e_{ij} &= h_{ij}^{e}v_d\ , \quad
m_Z = \frac{\sqrt{g^2 + g^{\prime 2}}v}{\sqrt{2}} \ , \quad
s_w = \frac{g'}{\sqrt{g^2 + g^{\prime 2}}} = \sqrt{1 - c_w^2}\ .
\end{align}

Given the Yukawa couplings $h^u_{ij}$, $h^d_{ij}$ and $h^e_{ij}$, the
Lagrangian (\ref{yukrpv}) predicts 108 R-parity breaking Yukawa couplings
in terms of 9 independent parameters which may be chosen as
\begin{align}
\m_i\ , \; B_i\ , \; m^2_{id} \quad \mathrm{or} \quad
\e_i\ , \; \e'_i \ , \; \e''_i \ .
\end{align}
These parameters determine lepton-gaugino mass mixings, lepton-slepton
and quark-slepton Yukawa couplings, and therefore the low-energy phenomenology.
The values of these parameters depend on the pattern of supersymmetry
breaking and the flavour structure of the supersymmetric standard model.

\section{Spontaneous R-parity breaking}

Let us now compute the parameters $\e_i$, $\e'_i$ and $\e''_i$
in a specific example where the spontaneous breaking
of R-parity is related to the spontaneous breaking of B-L, the difference of
baryon and lepton number \cite{bcx07}.

We consider a supersymmetric extension of the standard model with symmetry
group
\begin{align}
G = SU(3)\times SU(2)\times U(1)_Y \times U(1)_{B-L} \times U(1)_R \ .
\end{align}
In addition to
three quark lepton generations and the Higgs fields $H_u$ and $H_d$ the model
contains three right-handed neutrinos $\nu^c_i$, two non-Abelian singlets
$N^c$ and $N$, which transform as $\nu^c$ and its complex conjugate, respectively,
and three gauge singlets $X$, $\Phi$ and $Z$. The part of the superpotential
responsible for neutrino masses has the usual form
\begin{equation}\label{yuk}
W_{\n} = h_{ij}^{\n} l_i \n^c_j H_u
      +  \frac{1}{M_{\rm P}} h_{ij}^{n} \n^c_i\n^c_j N^2 \ ,
\end{equation}
where $M_{\rm P} = 2.4 \times 10^{18}\ {\rm GeV}$ is the Planck mass.
The expectation value of $H_u$ generates Dirac neutrino masses,
whereas the expectation value of the
singlet Higgs field $N$ generates the Majorana mass matrix of the right-handed
neutrinos $\n^c_i$. The superpotential responsible for B-L breaking is chosen as
\begin{equation}\label{bl}
W_{B-L} = X ( NN^c - \Phi^2)\ ,
\end{equation}
where unknown Yukawa couplings have been set equal to one.
$\Phi$ plays the role of a spectator field, which will finally be replaced
by its expectation value, $\langle \Phi \rangle = v_{B-L}$. Similarly,
$Z$ is a spectator field which breaks supersymmetry and $U(1)_R$,
$\langle Z \rangle = F_Z \theta\theta$. The superpotential
in Eqs.~(\ref{yuk}) and (\ref{bl}) is the most general one consistent
with the \mbox{R-charges} listed in Table~\ref{tab:Rcharge},
up to nonrenormalizable terms which are irrelevant for our discussion.

\begin{table}[b]
\begin{center}
\begin{tabular}{c|cccccccc}\hline \hline
$ $ & $\Psi$ & $H_u$ & $H_d$ & $N$ & $N^c$ & $\Phi$ & $X$ & $Z$  \\ \hline
$R$ & 1 & 0 & 0 & 0 & -2 & -1 & 4 & 0 \\ \hline\hline
\end{tabular}
\medskip
\caption{R-charges of matter fields $\Psi = q,u^c,e^c,d^c,l,\n^c$, Higgs fields and gauge singlets.}
\label{tab:Rcharge}
\end{center}
\end{table}

The expectation value of $\Phi$ leads to the breaking of $B-L$,
\begin{equation}\label{breaking}
\langle N \rangle = \langle N^c \rangle = \langle \Phi \rangle = v_{B-L}\;,
\end{equation}
where the first equality is a consequence of the $U(1)_{B-L}$ D-term.
This generates a Majorana mass matrix $M$ for the
right-handed neutrinos with three large eigenvalues $M_3 > M_2 > M_1$.
If the largest eigenvalue of $h^{n}$ is ${\cal O}(1)$, one has
$M_3 \simeq v_{B-L}^2/\Mp$. Integrating out the heavy Majorana neutrinos
one obtains the familiar dimension-5 seesaw operator which yields the light
neutrino masses.

Since the field $\Phi$ carries R-charge $-1$, the VEV $\langle \Phi \rangle$
breaks R-parity, which is conserved by the VEV $\langle Z \rangle$.
Thus, the breaking of $B-L$ is tied to the breaking of R-parity, which is then
transmitted to the low-energy degrees of freedom
via higher-dimensional operators in the superpotential and the K\"ahler
potential. Bilinear R-parity breaking, as discussed in the previous section,
is obtained from a correction to the K\"ahler potential,
\begin{align}\label{nonrem}
\Delta K = \frac{1}{\Mp^3} &\left(a_i Z^\dg \Phi^\dg N^c H_u l_i
         +  a'_i Z^\dg \Phi N^\dg H_u l_i \right)  \nonumber \\
         + \frac{1}{\Mp^4} &\left(b_i Z^\dg Z \Phi^\dg N^c H_u l_i
         +  b'_i Z^\dg Z\Phi N^\dg H_u l_i \right.  \nonumber \\
          &\left. + c_i Z^\dg Z \Phi^\dg N^c l_i^\dg H_d
         +  c'_i Z^\dg Z\Phi N^\dg l_i^\dg H_d \right)  + \mathrm{h.c.} \ .
\end{align}
Replacing the spectator fields $Z$ and $\Phi$, as well as $N^c$ and $N$ by
their expectation values, one obtains the correction to the superpotential
\begin{equation}\label{bilinear}
\Delta W =  \m_i H_u l_i \ , \nonumber
\end{equation}
with
\begin{equation}
\m_i = \sqrt{3}(a_i + a_i')\mG \Theta\ , \quad
\Theta = \frac{v_{B-L}^2}{\Mp^2} \simeq \frac{M_3}{\Mp}\;,
\label{Theta}
\end{equation}
where $\mG = F_Z/(\sqrt{3}\Mp)$ is the gravitino mass. Note that $\Theta$
can be increased or decreased by including appropriate Yukawa couplings
in Eqs.~(\ref{yuk}) and (\ref{bl}). The corresponding corrections to the
scalar potential are given by
\begin{equation}
 -\Delta {\cal L} = B_i H_u \tilde{l}_i
+ m^2_{id} {\tilde l}^\dagger_i H_d + \mathrm{h.c.} \ , \nonumber
\end{equation}
where
\begin{equation}\label{softTheta}
B_i = 3(b_i + b_i')m_{3/2}^2 \Theta\ , \quad
m^2_{id} = 3(c_i + c_i')m_{3/2}^2 \Theta\ .
\end{equation}

The corresponding R-parity conserving terms are generated by \cite{gm88}
\begin{align}\label{gm1}
K \supset \frac{a_0}{\Mp} Z^\dg H_u H_d
         + \frac{b_0}{\Mp^2} Z^\dg Z H_u H_d + \mathrm{h.c.} \ ,
\end{align}
which yields
\begin{align}\label{gm2}
W &\supset  \m H_u H_d \ , \hspace{1cm}
\m = \sqrt{3}a_0 m_{3/2} \ ,\\
-\mathcal{L} &\supset B H_u H_d + \mathrm{h.c.} \ ,\hspace{1cm}
B = 3 b_0 m_{3/2}^2 \ .
\end{align}

Higher dimensional operators yield further R-parity violating couplings between
scalars and fermions. However, the cubic couplings allowed by the symmetries
of our model
are suppressed by one power of $\Mp$ compared to ordinary Yukawa couplings
and cubic soft supersymmetry breaking terms. Note that the coefficients
of the nonrenormalizable operators are free parameters, which are only fixed
in specific models of supersymmetry breaking. In particular, one may have
$\mu^2, \widetilde{m}^2_i > m^2_{3/2}$ and hence a gravitino LSP.
All parameters are defined at the GUT scale and have to be evolved to the
electroweak scale by the renormalization group equations.

The phenomenological viability of the model depends on the size of
R-parity breaking mass mixings and therefore on the scale $v_{B-L}$
of R-parity breaking as well as the parameters $a_i \ldots c'_i$ in
Eq.~(\ref{nonrem}). Any model of flavour physics, which predicts Yukawa
couplings, will generically also predict the parameters $a_i \ldots c'_i$.
As a typical example, we use a model \cite{by99} for quark and lepton mass
hierarchies based on a Froggatt-Nielsen $U(1)$ flavour symmetry, which is
consistent with thermal leptogenesis and all contraints from flavour changing
processes \cite{bdh99}.

\begin{table}[b]
\begin{center}
\begin{tabular}{c|cccccccccccccc}\hline \hline
$\psi_i$ & $\T_3$ & $\T_2$ & $\T_1$ & $\F^*_3$ & $\F^*_2$ & $\F^*_1$ &
$\n^c_3$ & $\n^c_2$ & $\n^c_1$ & $H_u$ & $H_d$ & $\Phi$ & $X$ & $Z$ \\ \hline
$Q_i$ & 0 & 1 & 2 & 1 & 1 & 2 & 0 & 0 & 1 & 0 & 0 & 0 & 0 & 0 \\ \hline\hline
\end{tabular}
\medskip
\caption{Chiral $U(1)$ charges. $\T_i = (q_i,u^c_i,e^c_i)$, $\F = (d^c_i,l_i)$, $i=1\ldots 3$.}
\label{tab:chiralcharge}
\end{center}
\end{table}

The mass hierarchy is generated by the expectation
value of a singlet field $\phi$ with charge $Q_{\phi}=-1$ via
nonrenormalizable interactions with a scale
$\Lambda = \langle \phi\rangle/\eta > \Lambda_{GUT}$, $\eta \simeq 0.06$.
The $\eta$-dependence of Yukawa couplings and bilinear mixing
terms for multiplets $\psi_i$ with charges $Q_i$ is given by
\begin{equation}\label{mixingcharges}
h_{ij} \propto \eta^{Q_i + Q_j} , \quad \mu_i \propto \eta^{Q_i} , \quad
B_i \propto \eta^{Q_i} , \quad m^2_{id} \propto \eta^{Q_i}\ .
\end{equation}
The charges $Q_i$ for quarks, leptons, Higgs fields and singlets are listed
in Table~\ref{tab:chiralcharge}. The neutrino mass scale $m_\n \simeq 0.01$~eV implies for the
heaviest right-handed neutrinos $M_2 \sim M_3 \sim 10^{12}$~GeV. The
corresponding scales for $B-L$ breaking and R-parity breaking are
\begin{equation}\label{smallrpv}
v_{B-L} \simeq 10^{15}\ \mathrm{GeV}\ , \quad
\Theta = \frac{v_{B-L}^2}{\Mp^2} \simeq 10^{-6}\ .
\end{equation}
For the small R-parity breaking considered in this paper the neutrino masses
are dominated by the conventional seesaw contribution \cite{bcx07}.

The R-parity breaking parameters $\m_i$, $B_i$ and $m^2_{id}$ strongly depend
on the mechanism of supersymmetry breaking. In the example considered in this
section all mass parameters are $\mathcal{O}(m_{3/2})$, which corresponds
to gravity or gaugino mediation. From Eqs.~(\ref{Theta}),(\ref{softTheta})
and (\ref{mixingcharges}) one reads off
\begin{align}\label{param1}
\m_i = \hat{a}\eta^{Q_i} m_{3/2}\Theta\ , \quad
B_i = \hat{b}\eta^{Q_i} m^2_{3/2}\Theta\ , \quad
m^2_{id} = \hat{c}\eta^{Q_i} m^2_{3/2}\Theta\ ,
\end{align}
with $\hat{a},\hat{b},\hat{c} = \mathcal{O}(1)$. Correspondingly, one obtains
for $\epsilon$-parameters
(cf.~(\ref{rpv1newbasis}),(\ref{rpv2newbasis}))
\begin{align}\label{param2}
\e_i = a\eta^{Q_i} \Theta\ , \quad
\e'_i = b\eta^{Q_i} \Theta\ , \quad
\e''_{id} = c\eta^{Q_i} \Theta\ ,
\end{align}
with $a,b,c = \mathcal{O}(1)$. Our phenomenological analysis in Section~5.2
will be based on this parametrization of bilinear R-parity breaking.

Depending on the mechanism of supersymmetry breaking, the R-parity
breaking soft terms may vanish at the GUT scale \cite{add04},
\begin{align}
B_i(\Lambda_{\mathrm{GUT}}) = m^2_{id}(\Lambda_{\mathrm{GUT}}) = 0 \ .
\end{align}
Non-zero values of these parameters
at the electroweak scale are then induced by radiative corrections. The
renormalization group equations for the bilinear R-parity breaking mass terms
read (cf.~\cite{add04}, $t=\ln{\Lambda}$):
\begin{align}
    16 \pi^2 \frac{d \mu_i}{d t}
=&~ 3 \mu_i
    \left( h^u_{jk} h^{u*}_{jk}
         - \frac{1}{5} g_1^2
         - g_2^2 \right)
  + \mu_k h^e_{ij} h^{e*}_{kj}
  - \mu
    \left( \lambda_{ijk} h^{e*}_{kj}
         + 3 \lambda'_{kji} h^{d*}_{kj} \right) \ , \label{RG1} \\
    16 \pi^2 \frac{d B_i}{d t}
=&~ 3 B_i
    \left( h^u_{jk} h^{u*}_{jk}
         - \frac{1}{5} g_1^2
         - g_2^2 \right)
  + 6 \mu_i
    \left( \frac{1}{5} g_1^2 M_1
         + g_2^2 M_2 \right) \nonumber \\
 &+ B_k h^e_{ij} h^{e*}_{kj}
  - B
    \left( \lambda_{ijk} h^{e*}_{kj}
         + 3 \lambda'_{kji} h^{d*}_{kj} \right)\ , \label{RG2} \\
16 \pi^2 \frac{d m_{id}^2}{d t}
=&~ \lambda^*_{kji} h^e_{kj} m_d^2
  - m_{jd}^2 h^e_{jk} h^{e*}_{ik}
  - 3 \lambda'_{kji} h^d_{kj} m^2_d
  + h^e_{jk} h^{e*}_{jk} m_{id}^2 \nonumber \\
 &+ 3 h^{d*}_{kj} h^d_{kj} m_{id}^2
  + \widetilde m_{li}^2 \lambda^*_{nki} h^e_{nk}
  - 3 \widetilde m_{li}^2 \lambda'^*_{nki} h^e_{nk} \nonumber \\
 &+ 2 \lambda^*_{kji} \widetilde m_{lk}^2 \lambda_{kj}
  + 2 \lambda^*_{kji} h^e_{kj} \widetilde m_{ej}^2
  - 6 \lambda'^*_{kji} h^d_{kj} \widetilde m_{dk}^2
  - 6 \lambda'^*_{kji} \widetilde m_{qj}^2 h^d_{kj} \ .\label{RG3}
\end{align}
In bilinear R-parity breaking, the R-parity violating Yukawa couplings
vanish at the GUT scale. One-loop radiative corrections then yield
for the soft terms at the electroweak scale
(cf.~Eqs.~(\ref{RG1}),(\ref{RG2}); $\epsilon_i=\mu_i/\mu$):
\begin{align}\label{Birad}
B_i(\Lambda_{\mathrm{EW}}) = \frac{\mu_i}{16\pi^2}
\left(\frac{6}{5}g^{\prime 2} M_1 + 6 g^2 M_2\right)
\ln{\frac{\Lambda_{\mathrm{GUT}}}{\Lambda_{\mathrm{EW}}}}\ , \quad
m^2_{id}(\Lambda_{\mathrm{EW}}) = 0\ .
\end{align}
This illustrates that the bilinear R-parity breaking terms $\m^2_i$,
$B_i$ and $m^2_{id}$ are not necessarily of the same order of magnitude
at the electroweak scale.

\section{Neutral, charged and supercurrents}

In Section~2 we have discussed the R-parity breaking Yukawa couplings in
our model. For a phenomenological analysis we also need the couplings of
the gauge fields, i.e., photon, W-bosons and gravitino, to charged
and neutral matter,
\begin{align}
\mathcal{L} = - eJ_{e\mu}A^{\mu} - \frac{g}{c_w}J_{Z\mu}Z^{\mu}
              - \frac{g}{\sqrt{2}}J^-_{\mu}W^{+\mu}
              - \frac{g}{\sqrt{2}}J^+_{\mu}W^{-\mu}
              - \frac{1}{2\Mp}\overline{\psi}_{\mu} S^{\mu} \ .
\end{align}
The corresponding currents read
\begin{align}
	J_{e\mu} =&\overline{w}^+ \gamma_\mu {w}^+
		-\overline{w}^- \gamma_\mu {w}^-
		-\overline e_i \gamma_\mu e_i
                +\overline{e}_i^c \gamma_\mu e_i^c
  		-\overline{h}^-_d \gamma_\mu {h}^-_d
                +\overline{h}^+_u \gamma_\mu {h}^+_u \ , \label{egauge} \\
	J_{Z \mu} =&\overline{w}^+ \gamma_\mu w^+
		-\overline{w}^- \gamma_\mu w^-
		+\frac 12 \overline \nu_i \gamma_\mu \nu_i
		-\frac 12 \overline e_i \gamma_\mu e_i \nonumber \\
		&+\frac 12 \overline{h}^0_d \gamma_\mu h^0_d
		-\frac 12 \overline{h}^-_d \gamma_\mu h^-_d
		+\frac 12 \overline{h}^+_u \gamma_\mu h^+_u
		-\frac 12 \overline{h}^0_u \gamma_\mu h^0_u
		- s^2_w J_{e\mu} \ , \label{zgauge}\\
	J_\mu^- =&\sqrt 2 \left(\overline{w}^3 \gamma_\mu w^-
		-\overline{w}^+ \gamma_\mu w^3 \right)
		+ \overline \nu_i \gamma_\mu e_i
		+\overline{h}^0_d \gamma_\mu h^-_d
		+\overline{h}^+_u \gamma_\mu h^0_u \ , \label{wgauge} \\
        S^{\mu} =&\frac{i}{4}\left[\gamma^\nu,\gamma^\rho\right]\gamma^\mu
                 \left(\widetilde{b} B_{\nu\rho}
                 + \widetilde{w}^I W^I_{\nu\rho}\right) + \ldots .
                 \label{Ggauge}
\end{align}
The gravitino and the gauginos are now Majorana fermions,
\begin{align}
\widetilde{b} = b + b^c\ , \quad  \widetilde{w}^I = w^I + w^{cI}\ ,
\end{align}
where the superscript $c$ denotes charge conjugation.
In Eqs.~(\ref{egauge}) - (\ref{Ggauge}) we have only
listed contributions to the currents which will be
relevant in our phenomenological analysis.

The R-parity breaking described in the previous section leads to mass
mixings between the neutralinos $b$, $w^3$, $h_u^0$, $h_d^0$ with the neutrinos
$\n_i$, and the charginos $w^+$, $h_u^+$, $w^-$, $h_d^-$ with the charged
leptons $e^c_i$, $e_i$, respectively. The $7\times 7$ neutralino mass matrix
reads in the gauge eigenbasis
\begin{align}\label{Mneutral}
\mathcal{M}^N &=
\left(
\begin{array}{ccccc}
{M_1} & 0 & {m_Z} s_{\beta} s_w & -{m_Z} c_{\beta} s_w & - \zeta_i {m_Z} s_w \\
 0 & {M_2} & -{m_Z} s_{\beta} c_w & {m_Z} c_{\beta} c_w &  \zeta_i {m_Z} c_w \\
 {m_Z} s_{\beta} s_w & -{m_Z} s_{\beta}  c_w & 0 & -\mu  & 0 \\
 -{m_Z} c_{\beta} s_w & {m_Z} c_{\beta} c_w & -\mu  & 0 & 0 \\
 - \zeta_i {m_Z} s_w &  \zeta_i {m_Z}  c_w & 0 & 0 & 0
\end{array}
\right) \ ,
\end{align}
where we have neglected neutrino masses. Correspondingly, the $5\times 5$
chargino mass matrix which connects the states
$(w^-, h_d^-,e_i)$ and $(w^+,h_u^+,e^c_i)$ is given by
\begin{align}\label{Mcharged}
\mathcal M^C=&\left(
\begin{array}{ccccc}
 {M_2} & {m_Z} s_{\beta} c_w & 0 & 0 & 0 \  \\
 {m_Z} c_{\beta} c_w & \mu  & {\zeta_1}
 h^e_{11} \mu  & {\zeta_2} h^e_{22} \mu  & {\zeta_3} h^e_{33} \mu  \\
 {\zeta_1} {m_Z}  c_w & 0 & h^e_{11} v c_{\beta} & 0 & 0 \\
 {\zeta_2} {m_Z}   c_w & 0 & 0 & h^e_{22} v c_{\beta} & 0 \\
 {\zeta_3} {m_Z}   c_w & 0 & 0 & 0 & h^e_{33} v c_{\beta}
\end{array}
\right) \ .
\end{align}
Note that all gaugino and higgsino mixings with neutrinos and charged leptons
are parametrized by the three parameters $\zeta_i$.

\begin{figure}[t]
\begin{center}
\epsfig{file=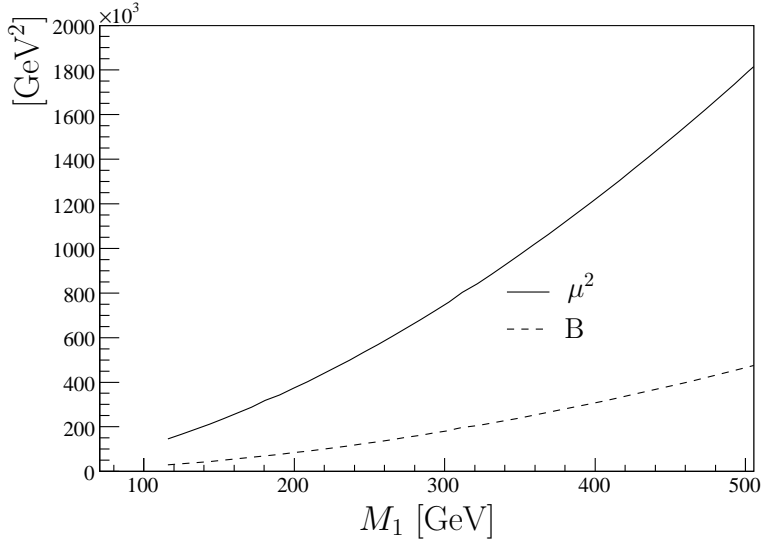,width=10cm}
\end{center}
\caption{The parameters $\mu$ and $B$ of Eqs.~(\ref{w1rp}) and
(\ref{scalarmass}), respectively, as functions of the bino mass $M_1$.
The plot has been obtained by means of \texttt{SOFTSUSY3.0} \cite{softsusy}.}
\label{fig:muB}
\end{figure}

In the following section we shall need the couplings of gravitino, $W$- and
$Z$-bosons to neutralino and chargino mass eigenstates. Since $\zeta_i \ll 1$,
diagonalization of the mass matrices to first order in $\zeta_i$ is obviously
sufficient. We shall also consider supergravity models
where the supersymmetry breaking parameters satisfy the inequalities
(cf.~Fig.~\ref{fig:muB})
\begin{align}\label{hierarchy}
m_Z < M_{1,2} < \m\ .
\end{align}
The gaugino-higgsino mixings are $\mathcal{O}(m_Z/\m)$, and therefore
suppressed, and $\chi_1^0$, the lightest neutralino, is bino-like.

The mass matrices $\mathcal{M}^N$ and $\mathcal{M}^C$ are diagonalized by
unitary and bi-unitary transformations, respectively,
\begin{align}
U^{(n)T}\mathcal{M}^N U^{(n)} =  \mathcal{M}^N_{\mathrm{diag}}\ ,\quad
U^{(c)\dagger}\mathcal{M}^C \widetilde{U}^{(c)} = \mathcal{M}^C_{\mathrm{diag}}\ ,
\end{align}
where $U^{(n)\dagger}U^{(n)} = U^{(c)\dagger}U^{(c)}
= \widetilde{U}^{(c)\dagger}\widetilde{U}^{(c)} = \1$. These unitary transformations
relate the neutral and charged gauge eigenstates to the mass eigenstates
$(\chi_a^0,\nu'_i)$ ($a=1,\ldots,4$) and $(\chi^-_\alpha, e'_i)$,
$(\chi^+_\alpha, e^{\prime c}_i)$ ($\alpha=1,2$), respectively. Inserting these
transformations in Eqs.~(\ref{zgauge}) - (\ref{Ggauge}) and dropping
prime superscripts, one obtains neutral, charged and supercurrents in the
mass eigenstate basis:
\begin{align}
J_{Z \mu} =& \overline \chi^0_a \gamma_\mu V^{(\chi^0)}_{ab} \chi^0_b
+\overline \chi^-_\alpha  \gamma_\mu V^{(\chi^-)}_{\alpha\beta} \chi^-_\beta
+\overline \chi^+_\alpha \gamma_\mu V^{(\chi^+)}_{\alpha\beta} \chi^+_\beta
+\overline \nu_i \gamma_\mu V^{(\nu)}_{ij} \nu_i
+\overline e_i \gamma_\mu V^{(e)}_{ij} e_i	\nonumber \\
&+\left(\overline \chi^0_a \gamma_\mu V^{(\chi,\nu)}_{a i} \nu_i
+\overline \chi^-_\alpha \gamma_\mu V^{(\chi^-,e)}_{\alpha i} e_i
+\overline \chi^+_\alpha \gamma_\mu V^{(\chi^+,e^c)}_{\alpha i} {e_i^c}
+{\rm h.c.} \right) \label{zmass} \ ,\\
J_{ \mu}^- =& \
\overline \chi^0_a \gamma_\mu V^{(\chi)}_{a\alpha} \chi^-_\alpha
+\overline \chi^0_a \gamma_\mu V^{(\chi,e)}_{ai} e_i
+\overline \nu_i \gamma_\mu V^{(\nu,\chi)}_{i\alpha} \chi^-_\alpha
+\overline \nu_i \gamma_\mu V^{(\nu,e)}_{ij} e_j \ , \label{wmass}\\
S^{\mu} = &\frac{i}{4}\left[\gamma^\nu,\gamma^\rho\right]\gamma^\mu
           \left(U^{(\widetilde\gamma,\chi)}_a \chi^0_a
           + U^{(\widetilde\gamma,\chi)*}_a \chi^{0c}_a
           + U^{(\widetilde\gamma,\nu)}_i\nu_i
           + U^{(\widetilde\gamma,\nu)*}_i\nu_i^c\right)
           F_{\nu\rho} + \ldots ,  \label{Gmass}
\end{align}
where we have defined the photino matrix elements
\begin{align}
U^{(\widetilde\gamma,\chi)}_a = c_w U^{(b,\chi)}_a + s_w U^{(w,\chi)}_a\ ,\quad
U^{(\widetilde\gamma,\nu)}_i = c_w U^{(b,\nu)}_i + s_w U^{(w,\nu)}_i\ .
\end{align}
In the appendix the unitary transformations between gauge and mass eigenstates
and the resulting matrix elements of neutral  and charged currents are given
to next-to-leading order in $m_Z/\mu$. As we shall see, that expansion
converges remarkably well.

In the next section we shall need the couplings of the lightest neutralino
$\chi^0_1$ to charged leptons and neutrinos, and the coupling of the gravitino
to photon and neutrino. From the formulae in appendix~A one easily
obtains\footnote{The matrix element $U_i^{(\tilde \gamma,\nu)}$ agrees with the
one used in \cite{it07,gr08} for $M_2-M_1\ll M_1$.}
($s_{2\beta} = 2 s_\beta c_\beta$)
\begin{align}
V_{1i}^{(\chi,\nu)} =& - \zeta_i \frac{m_Z s_w}{2 M_1}
\left( 1+\mathcal O \left(s_{2 \beta}\frac{m_Z^2}{\mu^2} \right) \right) \ ,
\label{meneutral} \\
V_{1i}^{(\chi,e)}=&- \zeta_i \frac{m_Z s_w}{ M_1}
\left( 1+\mathcal O \left(s_{2 \beta} \frac{m_Z^2}{\mu^2}\right)\right) \ ,
\label{mecharged} \\
U_i^{(\tilde \gamma,\nu)}=& \ \zeta_i \frac{m_Z s_w c_w
\left(M_2-M_1\right)}{M_1 M_2}
\left(1+\mathcal O \left(s_{2 \beta}\frac{m_Z^2}{\mu^2}\right)
\right) \ . \label{gravphoton}
\end{align}
Note that the charged and
neutral current couplings agree up to the isospin factor
at leading order in $m_Z^2/\mu^2$,
i.e., $V^{(\chi,\nu)}_{1i\ \text{LO}} = V^{(\chi,e)}_{1i\ \text{LO}}/2$.
The mass of the lightest neutralino is given by
\begin{align}\label{neutralinomass}
m_{\chi_1^0} = M_1 -
\frac{m_Z^2\left(M_1 + \mu s_{2\beta}\right)s_w^2}{\mu^2-M_1^2}
\left( 1+\mathcal O \left( \frac{ m_Z^2}{ \mu^2} \right) \right) \ .
\end{align}
We have numerically checked that varying $M_1$ between 120 and 500~GeV,
the relative corrections in Eqs.~(\ref{meneutral}) - (\ref{neutralinomass})
are less than 10\%.

\section{Fermi-LAT and the LHC}

We are now ready to evaluate the implications of recent Fermi-LAT data
\cite{FermiLAT1,FermiLAT2} and cosmological constraints \cite{cdx91,ehi09}
for signatures of decaying dark matter
at the LHC. We shall first discuss monochromatic gamma-rays produced by
gravitino decays and then analyze the implications for a neutralino and
a $\widetilde\tau$-NLSP, respectively.

\begin{figure}[t]
\begin{center}
\epsfig{file=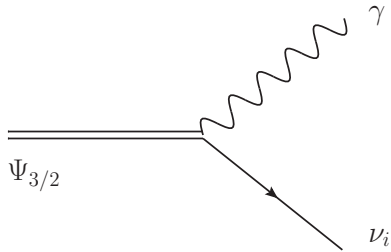,width=6cm}
\end{center}
\caption{Gravitino decay into photon and neutrino.}
\label{fig:gravitino}
\end{figure}

In order to keep our analysis transparent we shall not study the most general
parameter space of softly broken supersymmetry, but only consider two typical
boundary conditions for the supersymmetry breaking parameters of the MSSM
at the grand unification scale,
\begin{align}\label{modelA}
  (\text{A})~~~m_0 = m_{1/2},~~a_0 = 0,~~\tan\beta = 10\ ,
\end{align}
with equal universal scalar and gaugino masses, $m_0$ and $m_{1/2}$,
respectively; in this case a bino-like neutralino is the NLSP. The second
boundary condition corresponds to no-scale models or gaugino mediation,
\begin{align}\label{modelB}
  (\text{B})~~~m_0 = 0,~~m_{1/2},~~a_0 = 0,~~\tan\beta = 10 \ ,
\end{align}
which yields the right-handed stau as NLSP. In both cases, the trilinear
scalar coupling $a_0$ is put to zero for simplicity. Choosing
$\tan\beta = 10$ as a representative value of the Higgs vacuum expectation
values, only the gaugino mass parameter $m_{1/2}$ remains as
independent variable; the mass parameters $\mu$ and $B$ are determined by
requiring radiative electroweak symmetry breaking with the chosen ratio
$\tan\beta$.  For both boundary conditions (\ref{modelA}) and (\ref{modelB}),
the gaugino masses at the electroweak scale satisfy the familiar relations
\begin{align}\label{gauginomasses}
\frac{M_3}{M_1} \simeq 6.0\ , \quad \frac{M_2}{M_1} \simeq 1.9 \ .
\end{align}

For the chosen supergravity models, consistency with electroweak precision
tests, gravitino dark matter (GDM) and thermal leptogenesis leads to the
following
allowed mass ranges of gravitino and lightest neutralino \cite{bes08},
\begin{align}\label{massranges}
10~\mathrm{GeV} < m_{3/2} < 500~\mathrm{GeV}\ , \quad
100~\mathrm{GeV} < m_{\chi^0_1} < 500~\mathrm{GeV}\ ,
\end{align}
where we have used $m_{\chi^0_1} \simeq M_1$ (cf.~(\ref{neutralinomass})).
Note that the masses $M_1$ and $m_{3/2}$ cannot be chosen independently.
The GDM constraint implies that for a given gravitino mass
the maximal bino mass is
$M_1^{\mathrm{max}} \simeq 270~\mathrm{GeV}(m_{3/2}/100~\mathrm{GeV})^{1/2}$
\cite{bes08}.

Consider now the rate for gravitino decay into photon and neutrino\footnote{
$\Gamma_{3/2}(\gamma\nu)$ denotes the sum of the decay rates into photon
neutrino and photon antineutrino.}
\cite{ty00} (cf.~Fig.~\ref{fig:gravitino}),
\begin{align}\label{widthgravitino}
\Gamma_{3/2}(\gamma\nu) =
\frac{1}{32\pi} \sum_i|U^{(\widetilde\gamma, \nu)}_i|^2
\frac{m_{3/2}^3}{\Mp^2}\ ,
\end{align}
Inserting the matrix element (\ref{gravphoton}) one obtains the gravitino
lifetime
\begin{align}\label{gravlife}
\tau_{3/2}(\gamma \nu) =
\frac{32 \sqrt 2}{\alpha \zeta^2} \frac{G_F M_P^2}{m_{3/2}^3 }
\frac{M_1^2 M_2^2}{\left(M_2-M_1\right)^2}
\left( 1+\mathcal O \left( s_{2 \beta} \frac{m_Z^2}{\mu^2}\right)\right)\ ,
\end{align}
where $\alpha$ is the electromagnetic fine-structure constant, and we have
defined
\begin{align}
\zeta^2 = \sum_i \zeta_i^2 \ .
\end{align}
The corrections to the leading order expression in (\ref{gravlife}) are
less than 10\%.
Using Eq.~(\ref{gauginomasses}) and $\Mp = 2.4 \times 10^{18}$~GeV, one
obtains
\begin{align}\label{gravlife2}
\tau_{3/2}(\gamma\nu) =
1\times 10^{27} {\rm s}
\left(\frac{\zeta}{10^{-7}}\right)^{-2}
\left(\frac{M_1}{100~{\rm GeV}}\right)^{2}
\left(\frac{m_{3/2}}{10~\mathrm{GeV}}\right)^{-3}\ .
\end{align}

Recent Fermi-LAT data yield for dark matter decaying into 2 photons the
lower bound on the lifetime $\tau_{\rm DM}(\gamma\gamma) \gsim
1\times 10^{29}~\mathrm{s}$, which holds for photon energies in the range
$30~\mathrm{GeV} < E_\gamma < 200~\mathrm{GeV}$ \cite{FermiLAT1}.
For gravitino decays into photon and neutrino this implies
\begin{align}\label{fermilat}
\tau_{3/2}(\gamma\nu) \gsim 5\times 10^{28}~\mathrm{s} \ , \quad
30~\mathrm{GeV} < E_\gamma < 200~\mathrm{GeV} \ .
\end{align}
Since according to the GDM constraint the largest allowed bino mass
scales like $M_1^{\mathrm{max}} \propto m_{3/2}^{1/2}$, the largest lifetime
(\ref{gravlife2}), and therefore the most conservative bound on $\zeta$,
 is obtained for the smallest value of $m_{3/2}$.
For small gravitino masses, a rough lower bound on
the lifetime can be obtained from the isotropic diffuse gamma-ray flux.
The recent Fermi-LAT data give
$E^2 dJ/dE|_{5~\mathrm{GeV}}\simeq 3\times
10^{-7}~\mathrm{GeV}~(\mathrm{cm}^2~\mathrm{s}~\mathrm{str})^{-1}$
\cite{FermiLAT2}.
From the analysis in \cite{bbx07} one then obtains
 $\tau_{3/2} \gsim 10^{28}~\mathrm{s}$.\footnote{This lifetime is obtained by
rescaling in Fig.~2 of \cite{bbx07} the signal by the factor $0.1$.}
Together with Eq.~(\ref{gravlife2}) one then obtains the
approximate upper bound on the R-parity breaking parameter
\begin{align}\label{zetabound}
\zeta \lsim 3\times 10^{-8} \ .
\end{align}
On the other hand, the observation of a photon line corresponding to a
gravitino
lifetime close to the present bound would determine the parameter $\zeta$
as\footnote{The results (\ref{zetabound}) and (\ref{zetameasured}) are
approximately consistent with the recent analysis \cite{crx10}.}
\begin{align}\label{zetameasured}
\zeta_{\mathrm{obs}} = 10^{-9}
\left(\frac{5\times 10^{28}{\rm s}}{\tau_{3/2}(\gamma\nu)}\right)^{1/2}
\left(\frac{M_1}{200~{\rm GeV}}\right)
\left(\frac{m_{3/2}}{100~\mathrm{GeV}}\right)^{-3/2}\ .
\end{align}
Note the strong dependence of $\zeta_{\mathrm{obs}}$ on the gravitino mass.
In (\ref{zetameasured}) we have normalized these masses to
central values suggested by thermal leptogenesis, electroweak precision tests
and gravitino dark matter \cite{bes08}.

\subsection{Neutralino NLSP}

\begin{figure}[t]
\begin{center}
\epsfig{file=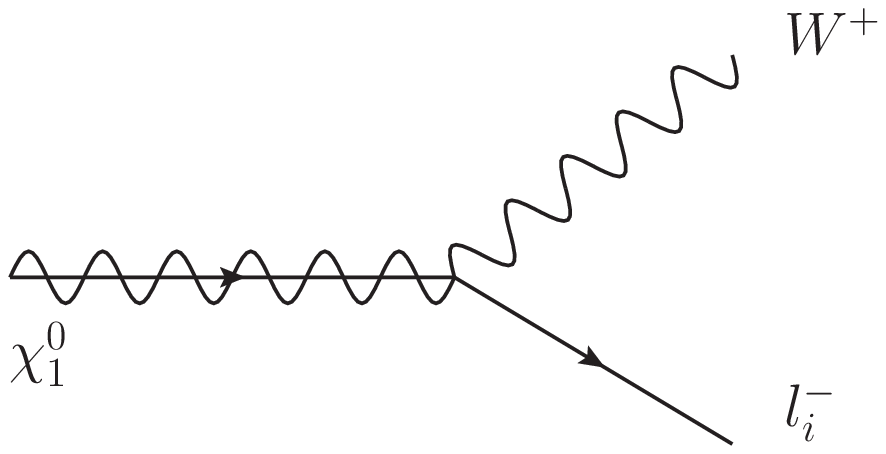,width=6cm}\hspace{2cm}
\epsfig{file=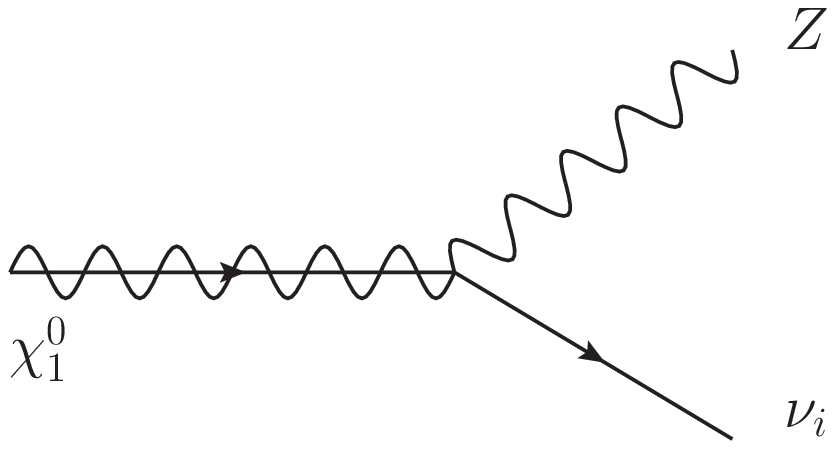,width=6cm}
\end{center}
\caption{Neutralino decays into charged lepton and W-boson, and neutrino and Z-boson.}
\label{fig:neutralino}
\end{figure}

\begin{figure}[t]
\begin{center}
\epsfig{file=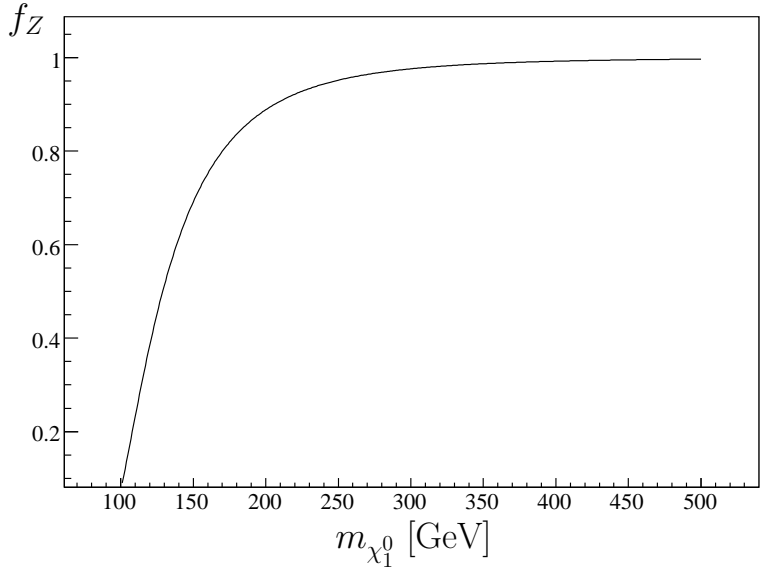,width=10cm}
\end{center}
\caption{Phase space suppression factor for neutralino decay to $Z$-boson
and neutrino.}
\label{fig:phasespace}
\end{figure}

A neutralino NLSP heavier than $100~\mathrm{GeV}$ dominantly decays into
charged lepton and W-boson or neutrino and Z-boson \cite{mrv98} (cf.~Fig.~\ref{fig:neutralino}).
The partial decay widths are given by
\begin{align}
\Gamma\left( \chi_1^0 \rightarrow W^\pm l^\mp \right) =&
\frac{G_F m_{\chi_1^0}^3}{4 \sqrt 2 \pi} \sum_i
\left| V_{1i \ \text{LO}}^{(\chi,e)} \right|^2 f_W (m_{\chi_1^0} )
\left(1+\mathcal O \left(s_{2\beta}\frac{m_Z^2}{\mu^2}\right)\right)\ ,\\
\Gamma\left( \chi_1^0 \rightarrow Z \nu \right) =&
\frac{G_F m_{\chi_1^0}^3}{2 \sqrt 2 \pi} \sum_i
\left| V_{1i\ \text{LO}}^{(\chi,\nu)} \right|^2 f_Z (m_{\chi_1^0})
\left( 1+\mathcal O \left(s_{2\beta}\frac{m_Z^2}{\mu^2}\right)\right)\ .
\end{align}
Here $V^{(\chi,e)}_{1i\ \text{LO}}$ and $V^{(\chi,\nu)}_{1i\ \text{LO}}$
are the charged and neutral current matrix elements at leading order,
which are given in Eqs.~(\ref{mecharged}) and (\ref{meneutral}), respectively,
and
\begin{align}
    f_{W,Z}(m_{\chi^0_1})
  = \left(1 - \frac{m_{W,Z}^2}{m_{\chi^0_1}^2}\right)^2
    \left(1 + 2 \frac{m_{W,Z}^2}{m_{\chi^0_1}^2}\right)
\end{align}
is a phase space factor which becomes important for neutralino masses close to
the lower bound for $m_{\chi^0_1}$ of $100$~GeV
(cf.~Fig.~\ref{fig:phasespace}).

The total neutralino NLSP width is the sum
\begin{align}
\Gamma_{\chi_1^0} = \Gamma(\chi_1^0 \rightarrow W^\pm l^\mp)
                    + \Gamma(\chi_1^0 \rightarrow Z \nu) \ .
\end{align}
Using the matrix elements (\ref{meneutral}) and (\ref{mecharged}),
one obtains the branching ratios
\begin{align}
BR\left(\chi_1^0 \rightarrow W^{\pm}l^{\mp}\right) \simeq
2\ BR\left(\chi_1^0 \rightarrow Z\nu\right) \ .
\end{align}
Furthermore, the flavour structure of our model implies
\begin{align}
BR\left(\chi_1^0 \rightarrow W^{\pm}\mu^{\mp}\right) \simeq
BR\left(\chi_1^0 \rightarrow W^{\pm}\tau^{\mp}\right) \ .
\end{align}

Using the matrix elements (\ref{meneutral}), (\ref{mecharged}) and
(\ref{gravphoton}) for neutral, charged and supercurrent, respectively,
one can express the neutralino lifetime directly in terms of the gravitino
lifetime,
\begin{align}\label{neutralinogravitino}
\tau_{\chi_1^0} =&\frac{c_w^2}{2 \sqrt 2} \frac{\left(M_2-M_1\right)^2}{M_2^2}
\frac{m_{3/2}^3}{G_F M_P^2 m_{\chi_1^0}^3} \tau_{3/2}(\gamma \nu) \nonumber \\
&\times\left(2 f_W(m_{\chi_1^0})+f_Z(m_{\chi_1^0})\right)^{-1}
\left( 1+\mathcal O \left(s_{2\beta}\frac{ m_Z^2}{ \mu^2} \right) \right) \ .
\end{align}
With the mass relations (\ref{neutralinomass}) and (\ref{gauginomasses}) one
then obtains for the minimal neutralino decay length
\begin{align}\label{minneutralino}
c \tau_{\chi_1^0} \gsim \ 80~\text{cm}
&\left(\frac{m_{\chi_1^0}}{150 \text{GeV}} \right)^{-3}
\left( \frac{m_{3/2}}{10 \text{GeV}} \right)^{3}
\left(\frac{\tau_{3/2}(\gamma\nu)}{1\times 10^{28}~\text{s}}\right)\nonumber \\
&\times\left(2 f_W(m_{\chi_1^0})+f_Z(m_{\chi_1^0})\right)^{-1}
\left( 1+\mathcal O \left(s_{2\beta}\frac{ m_Z^2}{ \mu^2} \right) \right) \ .
\end{align}
In Eqs.~(\ref{neutralinogravitino}) and (\ref{minneutralino}) the
corrections to the leading order expressions are less than 10\%.
We emphasize again the strong dependence of this lower bound on the neutralino
and gravitino masses. For instance, for a gravitino mass of $100~\mathrm{GeV}$
and the Fermi-LAT bound $\tau_{3/2} \gsim 5\times 10^{28}~\mathrm{s}$, which
applies for gravitino masses in the range
$60~\mathrm{GeV} < m_{3/2} < 400~\mathrm{GeV}$, one obtains
$c\tau_{\chi_1^0}\simeq 2~\mathrm{km}$ for a neutralino mass of
$150~\mathrm{GeV}$. It is very interesting that such neutralino lifetimes
are detectable at the LHC \cite{iim08}.

We conclude that, given the current bounds on the gravitino lifetime,
a neutralino NLSP may still decay into gauge boson and lepton inside the
detector, yielding a spectacular signature. However, for most of the parameter
space a neutralino NLSP decays outside the detector, leading to events
indistinguishable from ordinary neutralino dark matter.

\subsection{\boldmath $\widetilde\tau$-Lepton NLSP}

Contrary to the neutralino NLSP decay, the R-parity violating decays of
a $\widetilde\tau_1$-NLSP strongly depend on the flavour structure and
the supersymmetry breaking parameters. The relative strength of the various
decay modes becomes most transparent in the field basis where all bilinear
R-parity breaking terms vanish, as discussed in Section~2. Since the
R-parity breaking Yukawa couplings are proportional to the ordinary
Yukawa couplings, decays into fermions of the second and third generation
dominate. The leading partial decay widths of left- and right-handed
$\widetilde\tau$-leptons are
(cf.~(\ref{yukrpv}))
\begin{align}
\Gamma_{\widetilde\tau_L}(\tau_R\nu)
&= \frac{1}{16\pi}\sum_i|\hat{\lambda}_{i33}|^2 m_{\widetilde\tau_L}\ , \\
\Gamma_{\widetilde\tau_L}(\bar{t}_L b_R)
&= \Gamma_{\widetilde\tau_L}(\bar{t}_L s_R)
 = \frac{3}{16\pi}|\lambda'_{333}|^2 m_{\widetilde\tau_L}\ , \\
\Gamma_{\widetilde\tau_L}(\bar{t}_R b_L)
&= \frac{3}{16\pi}|\hat{\lambda}'_{333}|^2 m_{\widetilde\tau_L}\ , \\
\Gamma_{\widetilde\tau_R}(\tau_L\nu)
&= \Gamma_{\widetilde\tau_R}(\mu_L\nu)
 = \frac{1}{16\pi}\sum_i|\lambda_{i33}|^2 m_{\widetilde\tau_R}\ .
\end{align}
In the flavour model discussed in Section~3, the order of magnitude of the
various decay widths is determined by the power of the hierarchy parameter
$\eta$ ($\eta^2\simeq 1/300$),
\begin{align}
\Gamma_{\widetilde\tau_L}(\tau_R\nu)
&\sim \Gamma_{\widetilde\tau_R}(\tau_L\nu)
= \Gamma_{\widetilde\tau_R}(\mu_L\nu) \nonumber\\
&\sim \Gamma_{\widetilde\tau_L}(\bar{t}_L b_R)
\sim \Gamma_{\widetilde\tau_L}(\bar{t}_L s_R)
\sim \eta^4\Theta^2 m_{\widetilde\tau}\ , \\
\Gamma_{\widetilde\tau_L}(\bar{t}_R b_L)
&\sim \eta^2\Theta^2 m_{\widetilde\tau}\ .
\end{align}

\begin{figure}[t]
\begin{center}
\epsfig{file=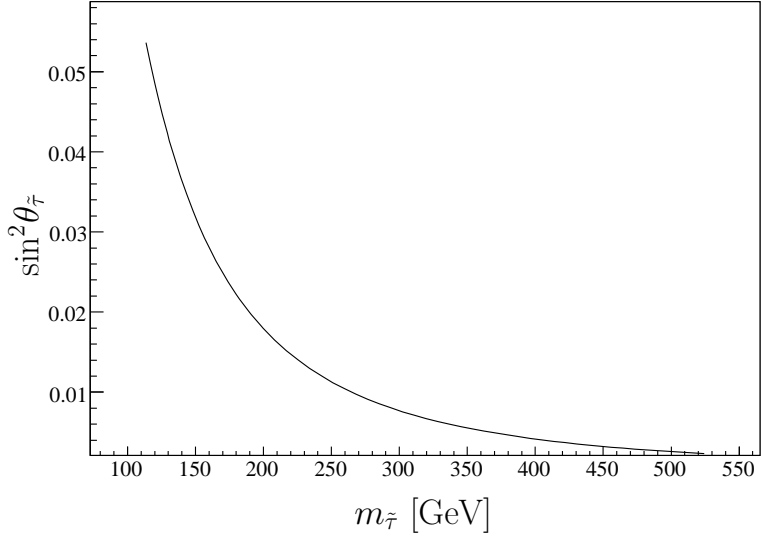,width=10cm}
\end{center}
\caption{$\widetilde\tau$-mixing angle: $\sin^2(\theta_\tau)$ as function
of the lightest $\widetilde\tau$-mass $m_{\widetilde\tau_1}$.}
\label{fig:stauangle}
\end{figure}

The lightest mass eigenstate $\widetilde\tau_1$ is a linear combination
of $\widetilde\tau_L$ and $\widetilde\tau_R$,
\begin{align}
\widetilde\tau_1 = \sin{\theta_\tau} \widetilde\tau_L
+ \cos{\theta_\tau} \widetilde\tau_R \ .
\end{align}
From the above equations one obtains the $\widetilde\tau_1$-decay width
\begin{align}
\Gamma_{\widetilde\tau_1}
= \sin^2{\theta_\tau}\left(\Gamma_{\widetilde\tau_L}(\tau_R\nu)
+ 2\Gamma_{\widetilde\tau_L}(\bar{t}_L b_R)
+ \Gamma_{\widetilde\tau_L}(\bar{t}_R b_L)\right)
+ 2 \cos^2{\theta_\tau}\Gamma_{\widetilde\tau_R}(\tau_L\nu) \ .
\end{align}
The total width is dominated by the contributions
$\widetilde\tau_R \rightarrow \tau_L\nu,\mu_L\nu$ and
$\widetilde\tau_L \rightarrow \bar{t}_R b_L$, respectively,
\begin{align}
\Gamma_{\widetilde\tau_1}
&= \sin^2{\theta_\tau} \Gamma_{\widetilde\tau_L}(\bar{t}_R b_L)
+ 2\cos^2{\theta_\tau}\Gamma_{\widetilde\tau_R}(\tau_L\nu) \ ,
\end{align}
and it can be directly expressed in
terms of the $\tau$-lepton and top-quark masses,
\begin{align}\label{true}
\Gamma_{\widetilde\tau_1} =
\frac{\epsilon^2}{16\pi v^2}
\left(3m_t^2\sin^2{\theta_\tau}
+ 2m_{\tau}^2 \tan^2{\beta}\cos^2{\theta_\tau}\right) m_{\widetilde\tau_1} \ ,
\end{align}
where we have assumed
\begin{align}
\epsilon_{2,3} = \epsilon'_{2,3} = \epsilon''_{2,3} \equiv \epsilon \ .
\end{align}
This corresponds to the parameter choice $a=b=c=1$ in Eq.~(\ref{param2}).
Note that $\widetilde\tau_1$-decay width and branching ratios have a
considerable uncertainty since these parameters depend on the unspecified
mechanism of supersymmetry breaking. From Eqs.~(\ref{zetabeta}), (\ref{Theta})
and $\eta\simeq 0.06$, one obtains for the R-parity breaking parameter
\begin{align}
\epsilon \simeq \zeta \simeq \eta\Theta \simeq 6\times 10^{-8} \ ,
\end{align}
which is consistent with the present upper bound (\ref{zetabound})
within the theoretical uncertainties.

The dependence of the mixing angle $\theta_\tau$ on $m_{\widetilde\tau_1}$
is shown in Fig.~\ref{fig:stauangle} for the boundary condition (\ref{modelB}).
For masses below the top-bottom threshold only leptonic
$\widetilde\tau_1$-decays are possible. When the decay into top-bottom
pairs becomes kinematically allowed, $\sin^2{\theta_\tau}$ is small.
However, the suppression by a small mixing angle is compensated by the
larger Yukawa coupling compared to the leptonic decay mode. This is a
direct consequence of the couplings $\hat{\lambda}'$ which were not taken into
account in previous analyses.

\begin{figure}
\begin{center}
\epsfig{file=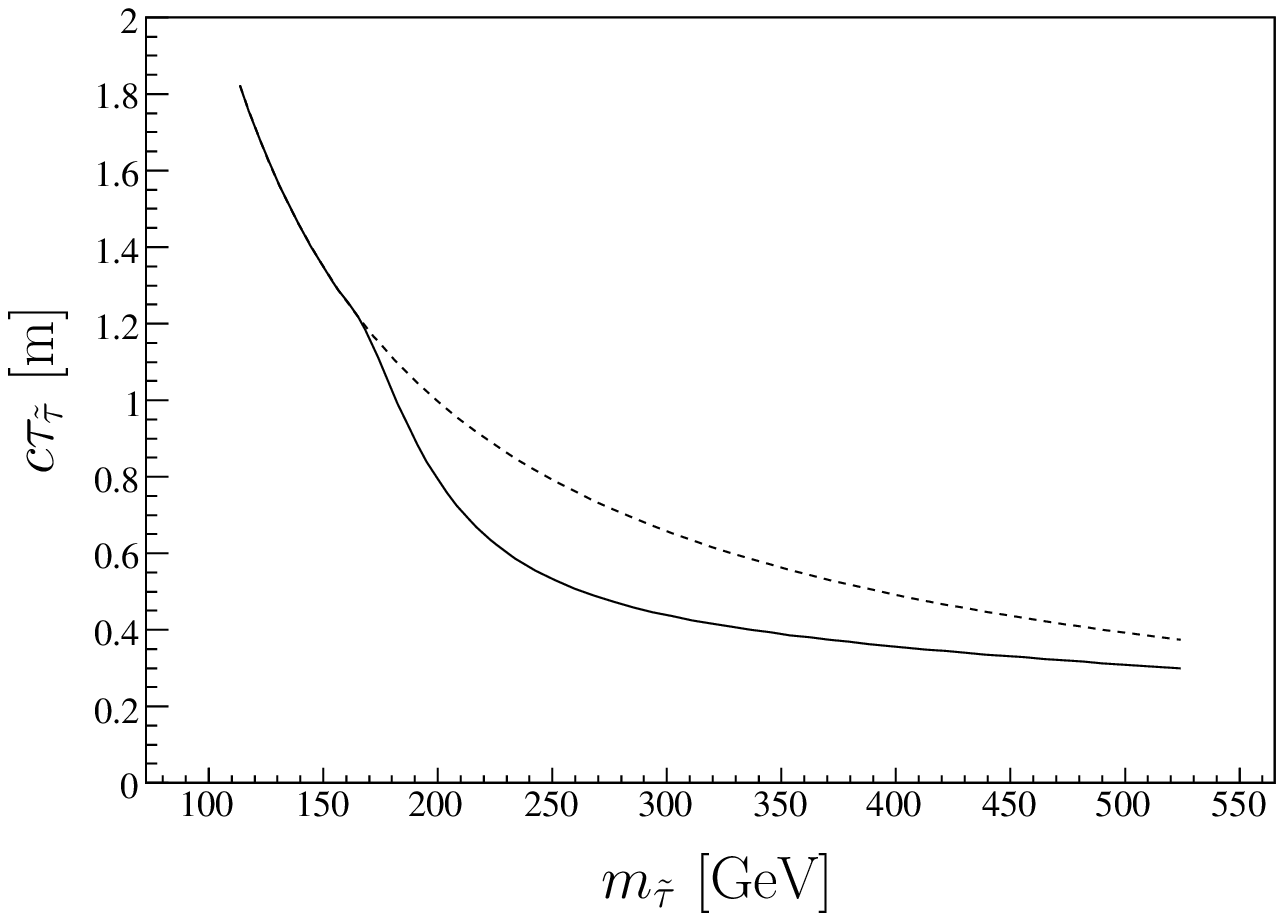,width=10cm}
\caption{$\widetilde\tau_1$-decay length as function of
$m_{\widetilde\tau_1}$. Above the top-bottom threshold hadronic decays
decrease the ${\widetilde\tau_1}$-lifetime.}
\label{fig:staudecaylength}
\vspace{2cm}
\epsfig{file=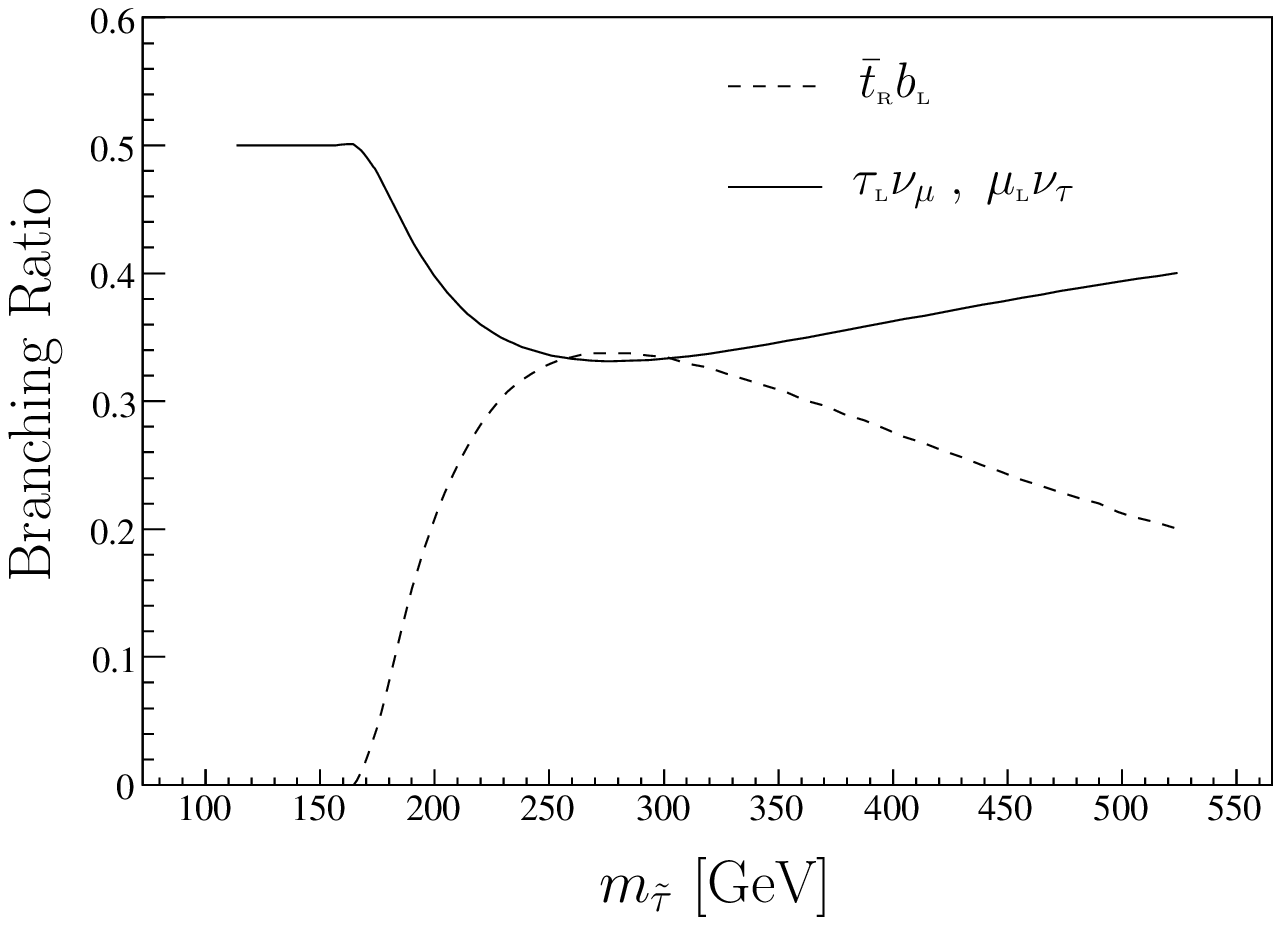,width=10cm}
\caption{$\widetilde\tau_1$-branching ratios as functions of
$m_{\widetilde\tau_1}$. The dependence on the ${\widetilde\tau_1}$-mass
is determined by the top-bottom threshold and the mass dependence of the
$\widetilde\tau_1$-mixing angle.}
\label{fig:branching}
\end{center}
\end{figure}

Due to the competition between mixing angle suppression and hierarchical
Yukawa couplings, the top-bottom threshold is clearly visible in the
$\widetilde\tau_1$-decay length as well as the branching ratios into leptons
and heavy quarks. This is illustrated in Figs.~\ref{fig:staudecaylength} and
\ref{fig:branching}, respectively, where these observables are plotted
as functions of $m_{\widetilde\tau_1}$. Representative values
of the $\widetilde\tau_1$-decay lengths below and above the top-bottom
threshold are
\begin{align}
m_{\widetilde\tau_1} < m_t + m_b: \quad
&c\tau_{\widetilde\tau_1}\big|_{150~\mathrm{GeV}} =
1.4~\mathrm{m}
\left(\frac{\epsilon}{5\times 10^{-8}}\right)^{-2}\ , \\
m_{\widetilde\tau_1} > m_t+m_b: \quad
&c\tau_{\widetilde\tau_1}\big|_{250~\mathrm{GeV}} =
0.6~\mathrm{m}
\left(\frac{\epsilon}{5\times 10^{-8}}\right)^{-2}\ .
\end{align}
Choosing for $\epsilon$ the representative value (\ref{zetameasured}) from
gravitino decay, $\epsilon = \zeta_{\mathrm{obs}} = 10^{-9}$, one obtains
$c\tau_{\widetilde\tau_1} = 4~\mathrm{km} (1~\mathrm{km})$ for
$m_{\widetilde\tau_1} = 150~\mathrm{GeV} (250~\mathrm{GeV})$.
It is remarkable that such lifetimes can be measured at the LHC
\cite{iim08,ahs09}.

Is it possible to avoid the severe constraint from gravitino decays on the
$\widetilde\tau_1$-decay length? In principle, both observables are
independent, and the unknown constants in the definition of $\epsilon$,
$\epsilon'$ and $\epsilon''$ can be adjusted such that $\zeta=0$.
However, this corresponds to a strong fine-tuning, unrelated to an
underlying symmetry. To illustrate this, consider the case where the
soft R-parity breaking parameters vanish at the GUT scale,
$B_i=m^2_{id}=0$, which was discussed in Section~3.
In bilinear R-parity breaking, also the R-parity violating Yukawa couplings
vanish at the GUT scale. With the one-loop radiative corrections at the
electroweak scale (cf.~(\ref{Birad}); $\epsilon_i=\mu_i/\mu$),
\begin{align}
B_i(\Lambda_{\mathrm{EW}}) = \frac{\epsilon_i\mu}{16\pi^2}
\left(\frac{6}{5}g^{\prime 2} M_1 + 6 g^2 M_2\right)
\ln{\frac{\Lambda_{\mathrm{GUT}}}{\Lambda_{\mathrm{EW}}}}\ , \quad
m^2_{id}(\Lambda_{\mathrm{EW}}) = 0\ , \nonumber
\end{align}
and $M_{1,2}\sim\mu$,
one reads off from Eqs.~(\ref{softnewbasis}), (\ref{rpv1newbasis})
and (\ref{rpv2newbasis})
\begin{align}
\epsilon'_i, \epsilon''_i = \mathcal{O}(\epsilon_i)\ .
\end{align}
Hence, all R-parity breaking parameters are naturally of the same order,
unless the fine-tuning also includes radiative corrections between the
GUT scale and the electroweak scale.

Even if one accepts the fine-tuning $\zeta=0$, one still has to satisfy the
cosmological bounds on R-parity violating couplings, which yield
$\epsilon_i=\mu_i/\mu \lsim 10^{-6}$ \cite{ehi09}. In the flavour model
discussed in Section~3 this corresponds to the choice $a=20$ in
Eq.~(\ref{param1}). For the smaller $\widetilde\tau_1$-mass, which is
preferred by electroweak precision tests, one then obtains the lower
bound on the decay length
\begin{align}\label{staubound}
c\tau_{\widetilde\tau_1}\big|_{150~\mathrm{GeV}} \gsim 4~\mathrm{mm}\ .
\end{align}
However, let us emphasize again that current constraints from Fermi-LAT on
the diffuse gamma-ray spectrum indicate decay lengths several orders of
magnitude larger.

\subsection{Planck Mass Measurement}

It has been pointed out in \cite{bcx07} that, in principle, one can
determine the Planck mass from decay properties of a $\widetilde\tau$-NLSP
together with the observation of a photon line in the diffuse gamma-ray flux,
which is produced by
gravitino decays. This is similar to the proposed microscopic determination of
the Planck mass based on decays of very long lived $\widetilde\tau$-NLSP's
in the case of a stable gravitino \cite{bhx04}.

From our analysis of NLSP decays in this section it is clear that neutralino
NLSP decays are particularly well suited for a measurement of the Planck
mass, which does not require any additional assumptions.
Eq.~(\ref{neutralinogravitino}) implies ($G_F = \sqrt{2}/(4v^2)$),
\begin{align}\label{mp1}
\Mp = &c_w v \frac{M_2-M_1}{M_2}
    \left(\frac{m_{3/2}}{m_{\chi^0_1}}\right)^{3/2}
    \left(\frac{\tau_{3/2}(\gamma\nu)}{\tau_{\chi^0_1}}\right)^{1/2}\nonumber\\
    &\times\left(2 f_{W}(m_{\chi^0_1})
    + f_{Z}(m_{\chi^0_1})\right)^{-1/2}
    \left(1 + \mathcal O \left(s_{2\beta}\frac{m_Z^2}{\mu^2} \right)\right)\ .
\end{align}
As expected, for gravitino and neutralino masses of the same order of
magnitude, the ratio of the two-body lifetimes is determined by the ratio
of the electroweak scale and the Planck mass,
\begin{align}
\frac{\tau_{\chi^0_1}}{\tau_{3/2}(\gamma\nu)} \sim \frac{v^2}{\Mp^2}\ .
\end{align}
Quantitatively, using the relation (\ref{gauginomasses}) for the gaugino
masses, one finally obtains ($v=174~\mathrm{GeV}$),
\begin{align}\label{mp2}
\Mp = &3.6\times 10^{18}~\mathrm{GeV}
    \left(\frac{m_{3/2}}{m_{\chi^0_1}}\right)^{3/2}
    \left(\frac{\tau_{3/2}(\gamma\nu)}{10^{28}~\mathrm{s}}\right)^{1/2}
    \left(\frac{\tau_{\chi^0_1}}{10^{-7}~\mathrm{s}}\right)^{-1/2}  \nonumber\\
    &\times\left(2 f_{W}(m_{\chi^0_1})
    + f_{Z}(m_{\chi^0_1})\right)^{-1/2}
    \left(1 + \mathcal O \left(s_{2\beta}\frac{m_Z^2}{\mu^2} \right)\right)\ .
\end{align}
It is remarkable that the observation of a photon line in the diffuse
gamma-ray flux, together with a measurement of the neutralino lifetime
at the LHC, can provide a microscopic determination of the Planck mass.

\section{Summary and conclusions}

We have studied a supersymmetric extension of the Standard Model with
small R-parity breaking related to spontaneous $B-L$ breaking, which is
consistent with primordial nucleosynthesis, thermal leptogenesis and
gravitino dark matter. We have considered supergravity models with
universal boundary conditions at the GUT scale, which lead to scalar tau
or bino-like neutralino as NLSP. Supersymmetry breaking terms have been
introduced by means of higher-dimensional operators. The size of the soft
terms corresponds to gravity or gaugino mediation.

We have analyzed our model, which represents a special case of bilinear
R-parity breaking, in a basis of scalar $SU(2)$ doublets,
where all bilinear terms vanish. In this basis one has R-parity
violating Yukawa and gaugino couplings. They are given in terms of ordinary
Yukawa couplings
and 9 R-parity breaking parameters $\epsilon_i$, $\epsilon'_i$ and
$\epsilon''_i$, $i=1,...,3$, which are constrained by the flavour symmetry
of the model. The R-parity violating couplings include terms
proportional to the up-quark Yukawa couplings, which
were not taken into accound in previous analyses.

The main goal of this paper are the quantitative connection between
gravitino decays and NLSP decays, and the corresponding implications of recent
Fermi-LAT data on the isotropic diffuse gamma-ray flux for superparticle
decays at the LHC. To
establish this connection one needs the relevant R-parity breaking matrix
elements of neutral, charged and supercurrents. For the considered
supergravity models these matrix elements can be obtained analytically
to good approximation, since the diagonalization of the neutralino-neutrino
and chargino-lepton mass matrices in powers of $m_Z/\mu$ converges well,
as demonstrated in the appendix.
The analytic expressions for the decay rates make the implications of the
Fermi-LAT data for NLSP decays very transparent.

Our main quantitative results are the branching ratios for NLSP decays and
the lower bounds on their decay lengths.
For a neutralino NLSP with $m_{\chi_1^0} = 150~\mathrm{GeV}$, the Fermi-LAT
data yield the lower bound $c\tau_{\chi_1^0} \gsim 30~\mathrm{cm}$.
This bound does not depend on details of the superparticle mass spectrum
or the flavour structure of the model. It directly follows from the
comparison of two-particle gravitino and neutralino decays.
On the contrary, there exists no model independent lower bound on the
$\widetilde\tau_1$-decay length. The natural relation between gravitino
and $\widetilde\tau$-decay widths can be avoided by fine-tuning. In
this case the cosmological constraint that the baryon asymmetry is not
washed out leads to the lower bound
$c\tau_{\widetilde\tau_1} \gsim 4~\mathrm{mm}$.

Without fine-tuning parameters the diffuse gamma-ray flux produced by
gravitino decays constrains the lifetime of a neutralino as well as a
$\widetilde\tau$-NLSP. For typical masses, $m_{3/2} \sim 100~\mathrm{GeV}$
and $m_{\mathrm{NLSP}} \sim 150~\mathrm{GeV}$,
the discovery of a photon line with an intensity
close to the present Fermi-LAT limit would imply
a decay length $c\tau_{\mathrm{NLSP}}$ of several hundered meters. This
is a definite prediction of a class of supergravity models. It is very
interesting that such lifetimes can be measured at the LHC
\cite{iim08,ahs09}.

Finally, it is intriguing that the observation of a photon line in the
diffuse gamma-ray flux, together with a measurement of the neutralino lifetime
at the LHC, can yield a microscopic determination of the Planck mass,
a crucial test of local supersymmetry.


\section*{Acknowledgements}

We would like to thank L.~Covi, M.~Grefe and C.~Weniger for helpful
discussions.
This work was supported by the German Science Foundation (DFG)
within the Collaborative Research Center 676 "Particles, Strings and
the Early Universe" and the Hamburg Excellence Initiative ``Connecting
Particles with the Cosmos''.


\begin{appendix}

\numberwithin{equation}{section}

\section{Appendix: Gauge and mass eigenstates}

\subsection{Mass matrix diagonalization}

The mass matrices $\mathcal M^N$ and $\mathcal M^C$ in the gauge eigenbasis were
explicitly given in Eqs.~(\ref{Mneutral}) and (\ref{Mcharged}), respectively,
\begin{align}
\mathcal{M}^N &=
\left(
\begin{array}{ccccc}
{M_1} & 0 & {m_Z} s_{\beta} s_w & -{m_Z} c_{\beta} s_w & - \zeta_i {m_Z} s_w \\
 0 & {M_2} & -{m_Z} s_{\beta} c_w & {m_Z} c_{\beta} c_w &  \zeta_i {m_Z} c_w \\
 {m_Z} s_{\beta} s_w & -{m_Z} s_{\beta} c_w & 0 & -\mu  & 0 \\
 -{m_Z} c_{\beta} s_w & {m_Z} c_{\beta} c_w & -\mu  & 0 & 0 \\
 - \zeta_i {m_Z}  s_w & \zeta_i {m_Z} c_w & 0 & 0 & 0
\end{array}
\right) \ , \nonumber \\
\mathcal M^C &=\left(
\begin{array}{ccccc}
 {M_2} & {m_Z} s_{\beta} c_w & 0 & 0 & 0 \  \\
 {m_Z} c_{\beta} c_w & \mu  & {\zeta_1}
 h^e_{11} \mu  & {\zeta_2} h^e_{22} \mu  & {\zeta_3} h^e_{33} \mu  \\
 {\zeta_1} {m_Z}  c_w & 0 & h^e_{11} v c_{\beta} & 0 & 0 \\
 {\zeta_2} {m_Z}   c_w & 0 & 0 & h^e_{22} v c_{\beta} & 0 \\
 {\zeta_3} {m_Z}   c_w & 0 & 0 & 0 & h^e_{33} v c_{\beta}
\end{array}
\right) \ . \nonumber
\end{align}
For non-vanishing $R$-parity breaking parameters $\zeta_i$, $i=1,\ldots,3$, they
 induce a mixing between gauginos, Higgsinos and leptons,
\begin{align}
- \mathcal L \supset &\frac 12 \left( b, w^3,h_u^0,h_d^0, \nu_i \right)
\mathcal M^N \left( b, w^3,h_u^0,h_d^0, \nu_i \right)^T \nonumber\\
&+\left(\left(w^-,h_d^-,e_i\right) \mathcal M^C \left(w^+,h_u^+,e_i^c\right)^T
+ \mathrm{h.c.}\right)\ .
\end{align}
The matrices $\mathcal M^N$ and $\mathcal M^C$ are diagonalized by unitary and
bi-unitary transformations, respectively,
\begin{align}
\label{eq:defU}
U^{(n)T}\mathcal{M}^N U^{(n)} =  \mathcal{M}^N_{\mathrm{diag}}\ ,\quad
U^{(c)\dagger}\mathcal{M}^C \widetilde{U}^{(c)} = \mathcal{M}^C_{\mathrm{diag}}\ ,
\end{align}
where $U^{(n)\dagger}U^{(n)} = U^{(c)\dagger}U^{(c)}
= \widetilde{U}^{(c)\dagger}\widetilde{U}^{(c)} = \1$. These unitary transformations
relate the neutral and charged gauge eigenstates to the mass eigenstates
$(\chi_a^0,\nu'_i)$ ($a=1,\ldots,4$) and $(\chi^-_\alpha, e'_i)$,
$(\chi^+_\alpha, e^{\prime c}_i)$ ($\alpha=1,2$), respectively.

In this work we consider the two boundary conditions $(A)$ and $(B)$, defined in
Eqs.~\eqref{modelA} and \eqref{modelB}, respectively. The corresponding supergravity
 models satisfy the relation
(\ref{hierarchy}), $m_Z < M_{1,2} < \mu$, and in the regime
$120 \ \text{GeV}\lesssim M_1 \lesssim 500 \ \text{GeV}$
one finds $0.07 \lesssim m_Z/\mu \lesssim 0.25$. We diagonalized the above
mass matrices to first order in the small parameters $\zeta_i$ and to second
order in $m_Z/\mu$. The size of the relative corrections given below has
been calculated for the above parameter range using \texttt{SOFTSUSY3.0}
\cite{softsusy}.
As we shall see, the relative corrections are of order $m_Z^2/\mu^2$, and
the expansion converges well for most matrix elements.

The neutralino and neutrino mass eigenvalues are
\begin{align}
m_{\chi^0_1} =& \ M_1 -
\frac{m_Z^2\left(M_1 + \mu s_{2\beta}\right)s_w^2}{\mu^2-M_1^2}
\left( 1+\mathcal O \left(\frac{ m_Z^2}{ \mu^2} \right) \right)\ , \\
m_{\chi^0_2} =& \ M_2 -
\frac{m_Z^2\left(M_2 + \mu s_{2\beta} \right)c_w^2}{\mu^2-M_2^2}
\left( 1+\mathcal O \left(\frac{ m_Z^2}{ \mu^2}\right)\right) \ , \\
m_{\chi^0_3} =& \ \mu + \frac{m_Z^2 \left(\mu-M_1 c_w^2-M_2 s_w^2\right)
\left(1 + s_{2\beta}\right)}{2 \left(\mu-M_1 \right) \left(\mu-M_2\right)}
\left(1 +\mathcal O\left( \frac{m_Z^2}{\mu^2} \right)\right) \ , \\
m_{\chi^0_4} =&-\mu - \frac{m_Z^2 \left(\mu+M_1 c_w^2+M_2 s_w^2\right)
\left(1- s_{2 \beta} \right)}{2 \left(\mu+M_1 \right) \left(\mu+M_2\right)}
\left(1 + \mathcal O\left(s_{2\beta}\frac{m_Z^2}{\mu^2} \right)\right) \ , \\
m_{\nu_i} =& \ 0+\mathcal O\left(\zeta^2\frac{m_Z^2 }{\mu^2}  \right) \ .
\end{align}
We checked numerically that relative corrections $\mathcal O (m_Z^2/\mu^2)$ to
the above neutralino masses are smaller than $0.05, 0.15, 0.10, 0.001$, for
$m_{\chi_1^0}, \ldots, m_{\chi_4^0}$, respectively.

The chargino and lepton mass eigenvalues are
\begin{align}
m_{\chi_1^\pm}=& \ M_2 - \frac{{m_Z}^2 ({M_2} + \mu  s_{2\beta})c_w^2}{2
\left(\mu ^2-{M_2}^2\right)}\left(1
+\mathcal O\left( \frac{m_Z^2}{\mu^2}\right) \right)\ , \\
m_{\chi_2^\pm}=& \ \mu  + \frac{{m_Z}^2 (\mu + M_2 s_{2\beta})c_w^2}{2
\left(\mu ^2-{M_2}^2\right)} \left(1
+\mathcal O\left(\frac{m_Z^2}{\mu^2}\right) \right)\ , \\
m_{e_i'}=& \ h^e_{ii} v c_\beta
\left(1 +\mathcal O\left(\zeta^2 \frac{m_Z^2}{\mu^2}\right)\right) \ .
\end{align}
Here the relative corrections of $\mathcal O (m_Z^2/\mu^2)$ are numerically smaller
than $5 \%$.

The unitary matrix $U^{(n)}$ from Eq.~\eqref{eq:defU} can be written as
\begin{align}
U^{(n)} =\left(\begin{array}{c|c}
\jvs U_{ab}^{(\chi^0)} & U_{ai}^{(\chi^0, \nu)} \\ \hline
\jvs U_{ia}^{(\nu, \chi^0)} & U_{ij}^{(\nu)}
\end{array}\right) \ ,
\end{align}
with
\begin{align}
\label{eq:Uchi}
U_{ab}^{(\chi^0)}=&\left(
\begin{array}{llll}
1 & 0 & 0 & 0 \\
0 & 1 & 0 & 0 \\
0 & 0 & -\frac{1}{\sqrt2} & \frac{1}{\sqrt2} \\
0 & 0 & \frac{1}{\sqrt2} & \frac{1}{\sqrt2}
\end{array}\right) \nonumber \\&\hskip -1.3cm +
\left(
\begin{array}{llll}
 -\frac{{m_Z}^2 \left({M_1}^2+2 \mu  s_{2\beta} {M_1}+\mu
^2\right) s_w^2}{2 \left({M_1}^2-\mu ^2\right)^2} &
\frac{{m_Z}^2 ({M_2}+\mu  s_{2\beta}) s_{2w}}{2 ({M_1}-{M_2})
\left({M_2}^2-\mu ^2\right)} & \frac{{m_Z} (c_\beta+s_\beta)
s_w}{\sqrt{2} ({M_1}-\mu )} & \frac{{m_Z} (c_\beta-s_\beta)
s_w}{\sqrt{2} ({M_1}+\mu )} \\
 -\frac{{m_Z}^2 ({M_1}+\mu  s_{2\beta}) s_{2w}}{2 ({M_1}-{M_2})
\left({M_1}^2-\mu ^2\right)} & -\frac{{m_Z}^2 c_w^2
\left({M_2}^2+2 \mu  s_{2\beta} {M_2}+\mu ^2\right)}{2
\left({M_2}^2-\mu ^2\right)^2} & -\frac{{m_Z} c_w
(c_\beta+s_\beta)}{\sqrt{2} ({M_2}-\mu )} & \frac{{m_Z} c_w
(s_\beta-c_\beta)}{\sqrt{2} ({M_2}+\mu )} \\
 \frac{{m_Z} (\mu  c_\beta+{M_1} s_\beta) s_w}{{M_1}^2-\mu ^2} &
-\frac{{m_Z} c_w (\mu  c_\beta+{M_2} s_\beta)}{{M_2}^2-\mu ^2} &
 \frac{{m_Z}^2  (c_\beta+s_\beta)}{
\mu ^2}x_1 &
\frac{ (c_\beta-s_\beta) {m_Z}^2}{ \mu ^2} x_2 \\
 -\frac{{m_Z} ({M_1} c_\beta+\mu  s_\beta) s_w}{{M_1}^2-\mu ^2}
& \frac{{m_Z} c_w ({M_2} c_\beta+\mu  s_\beta)}{{M_2}^2-\mu ^2}
& \frac{ (c_\beta+s_\beta) {m_Z}^2}{ \mu ^2} x_3 &
 \frac{ (c_\beta-s_\beta) {m_Z}^2}{ \mu ^2} x_4
\end{array}
\right) \nonumber \\
& \times \left(1+\mathcal O \left(\frac{m_Z^2}{\mu^2}\right)\right)
\ ,
\end{align}
where we used the abbreviations
\begin{align}
x_1 =&\ \frac{\mu}{4 \sqrt{2}}  \left(\frac{({M_2} s_\beta-({M_2}-2 \mu ) c_\beta)
c_w^2}{({M_2}-\mu )^2}+\frac{({M_1} s_\beta-({M_1}-2 \mu )
c_\beta) s_w^2}{({M_1}-\mu )^2}\right) \ , \\
x_2 =& \ \frac{\mu}{4 \sqrt{2}}   \left(-\frac{(({M_2}+2 \mu ) c_\beta+{M_2} s_\beta)
c_w^2}{({M_2}+\mu )^2}-\frac{(({M_1}+2 \mu ) c_\beta+{M_1}
s_\beta) s_w^2}{({M_1}+\mu )^2}\right) \ , \\
x_3 =& \ \frac{\mu}{4 \sqrt{2}}   \left(\frac{(({M_2}-2 \mu ) s_\beta-{M_2} c_\beta)
c_w^2}{({M_2}-\mu )^2}+\frac{(({M_1}-2 \mu ) s_\beta-{M_1}
c_\beta) s_w^2}{({M_1}-\mu )^2}\right) \ , \\
x_4 =& \ \frac{\mu}{4 \sqrt{2}}   \left(\frac{({M_2} c_\beta+({M_2}+2 \mu ) s_\beta)
c_w^2}{({M_2}+\mu )^2}+\frac{({M_1} c_\beta+({M_1}+2 \mu )
s_\beta) s_w^2}{({M_1}+\mu )^2}\right) \ .
\end{align}
The numerical error of the matrix \eqref{eq:Uchi} in our parameter range of interest
is smaller than $40 \%$ of the given NLO term. We do not discuss the
slow convergence for this R-parity conserving sub-matrix further, since this
is beyond the scope of our analysis.

Furthermore,
\begin{align}
\label{eq:Uchi0nu}
U_{ai}^{(\chi^0, \nu)}=& \ \zeta_i
\left(
\begin{array}{c}
 s_w  \frac{  {m_Z} }{{M_1} } \\
 - c_w\frac{  {m_Z}  }{{M_2} } \\
 -\frac{  {m_Z}^2  c_\beta \left({M_1} c_w^2+{M_2}
s_w^2\right)}{{M_1} {M_2} \mu } \\
 \frac{  {m_Z}^2  s_\beta \left({M_1} c_w^2+{M_2}
s_w^2\right)}{{M_1} {M_2} \mu }
\end{array}
\right)\left(1+\mathcal O \left(
s_{2\beta} \frac{m_Z^2}{\mu^2} \right)\right)\ , \\
U_{ia}^{(\nu, \chi^0)}=& \ \zeta_i
\left(
\begin{array}{c}
 -s_w\frac{  {m_Z}  }{{M_1}} \\
 c_w\frac{  {m_Z}  }{{M_2}} \\
 \frac{  {m_Z}^2  \left({M_1}c^2_w+{M_2}s_w^2- \mu \right)
 (c_\beta+s_\beta)}{ \sqrt{2} ({M_1}-\mu ) \mu  (
\mu -{M_2})} \\
 \frac{  {m_Z}^2  \left({M_1}c_w^2+{M_2}s_w^2+ \mu \right) (c_\beta-s_\beta)}{ \sqrt{2} \mu  ({M_1}+\mu )
({M_2}+\mu )}
\end{array}
\right)\left(1 + \mathcal O \left(\frac{m_Z^2}{\mu^2}\right)\right) \ , \\
U_{ij}^{(\nu)}=& \ \delta_{ij} +\mathcal O \left( \zeta^2 \frac{m_Z^2}{\mu^2}
\right) \ .
\end{align}
The uncertainties in Eq.~\eqref{eq:Uchi0nu} evaluate numerically to less than $5\%$.
For $U_{ia}^{(\nu, \chi^0)}$ they are less than $0.15, 0.10,0.25, 0.25$, for $a=1,\ldots,4$,
respectively.

The unitary matrices $U^{(c)}$ and $\widetilde U^{(c)}$ which diagonalize the matrix $\mathcal M^C$,
cf.~Eq.~\eqref{eq:defU}, can be denoted as
\begin{align}
\widetilde U^{(c)} &=\left(\begin{array}{c|c}
\jvs \widetilde U_{\alpha\beta}^{(\chi^+)} &
\widetilde U_{\alpha i}^{(\chi^+, e^c)} \\
\hline \jvs \widetilde U_{i \alpha}^{(e^c, \chi^+)} &
\widetilde U_{ij}^{(e^c)}
\end{array}\right) \ , &
 U^{(c)} &=\left(\begin{array}{c|c}
\jvs  U_{\alpha\beta}^{(\chi^-)} &
 U_{\alpha i}^{(\chi^-, e)} \\ \hline \jvs U_{i \alpha}^{(e, \chi^-)} &
 U_{ij}^{(e)}
\end{array}\right) \ .
\end{align}
We find
\begin{align}
\widetilde U_{\alpha\beta}^{(\chi^+)} =&
\left(
\begin{array}{cc} 1 & 0 \\ 0 & 1
\end{array}
\right)+
\left(
\begin{array}{cc}
 -\frac{{m_Z}^2 c_w^2 (\mu  c_\beta+{M_2} s_\beta)^2}{2
\left({M_2}^2-\mu ^2\right)^2} & -\frac{{m_Z} c_w (\mu  c_\beta+{M_2}
s_\beta)}{{M_2}^2-\mu ^2} \\
 \frac{{m_Z} c_w (\mu  c_\beta+{M_2} s_\beta)}{{M_2}^2-\mu ^2} &
-\frac{{m_Z}^2 c_w^2 (\mu  c_\beta+{M_2} s_\beta)^2}{2
\left({M_2}^2-\mu ^2\right)^2}
\end{array}
\right) \left(1+\mathcal O \left(
\frac{m_Z^2}{\mu^2}
\right)\right) \ , \\
\widetilde U_{ij}^{(e^c)} =& \ \delta_{ij}+\mathcal O \left( \zeta^2 \right) \ .
\end{align}
Numerically, the relative correction to the NLO contribution to $\widetilde U_{\alpha\beta}^{(\chi^+)}$ is less than $25\%$.
The off-diagonal elements of the matrix $\widetilde U^{(c)}$
to leading order in $h_{ii}^e$ are
\begin{align}
\widetilde U_{\alpha i}^{(\chi^+, e^c)} =& -\zeta_i {h_{ii}^e}
\left( \begin{array}{c} 0 \\ 1 \end{array} \right)
\nonumber \\&+ \zeta_i h_{ii}^e
\left(
\begin{array}{c}
 \frac{ {m_Z}  c_w ({M_2} s_\beta-v
c_\beta)}{{M_2}^2} \\
 \frac{{m_Z}^2 c_\beta c_w^2 (v \mu
c_\beta+{M_2} (v-\mu ) s_\beta)}{{M_2}^2 \mu ^2}
\end{array}
\right)  \left(1 +\mathcal O \left(\begin{array}{c}
\frac{s_{2 \beta} m_Z^2}{\mu^2}\\
\frac{ m_Z^2}{\mu^2}
\end{array}\right)\right)\ , \\
\widetilde U_{i \alpha}^{(e^c, \chi^+)} =& \ \zeta_i {{h_{ii}^e}}
\left(
\begin{array}{c}
 0 \\
 1
\end{array}
\right)
\nonumber \\&+ \zeta_i {{h_{ii}^e}}
\left(
\begin{array}{c}
 \frac{{m_Z}  c_w \left({M_2} s_\beta \mu ^2+
\left({M_2}^2 (v+\mu )-v \mu ^2\right)
c_\beta\right)}{{M_2}^2 \left(M_2^2- \mu ^2\right)} \\
 -\frac{ {m_Z}^2 \mu^2 c_w^2 y}{2 \left(\mu ^2-{M_2}^2
\right)^2}
\end{array}
\right)
\left(1 +\mathcal O \left(
\frac{ m_Z^2}{\mu^2}
\right)\right)\ ,
\end{align}
where
\begin{align}
y =& \frac{1}{\mu^4}\left(v s_{2\beta}
{M_2}^3+\mu  \left(2 v {M_2}^2+\mu ^2 (\mu -2 v)\right)
{c_\beta^2}+\mu  s_\beta \left(\mu  \left(2 \mu
^2-{M_2}^2\right) s_\beta
\right. \right. \nonumber \\ &\left. \left.
-2 {M_2} \left({M_2}^2+(v-2 \mu ) \mu
\right) c_\beta\right)\right)\  .
\end{align}
The numerical relative correction to the NLO term in
$\widetilde U_{i \alpha}^{(e^c, \chi^+)}$ is smaller than $0.10, 0.15$ for
$\alpha=1,2$, respectively.
For $\widetilde U_{1i}^{(\chi^+, e^c)}$
it is smaller\footnote{The numerical calculation of the error reaches
our numerical precision. The given value is calculated from the
comparison with the
analytical NNLO expression.} than $1\%$, and smaller than $10 \%$ for $\widetilde U_{2i}^{(\chi^+, e^c)}$.

The block diagonal elements of the matrix $U^{(c)}$ are
\begin{align}
 U_{\alpha\beta}^{(\chi^-)} =&
\left(
\begin{array}{cc} 1 & 0 \\ 0 & 1
\end{array}
\right)+
 \left(
\begin{array}{ll}
 -\frac{{m_Z}^2 c_w^2 ({M_2} c_\beta+\mu  s_\beta)^2}{2
\left({M_2}^2-\mu ^2\right)^2} & -\frac{{m_Z} c_w ({M_2}
c_\beta+\mu  s_\beta)}{{M_2}^2-\mu ^2} \\
 \frac{{m_Z} c_w ({M_2} c_\beta+\mu  s_\beta)}{{M_2}^2-\mu ^2} &
-\frac{{m_Z}^2 c_w^2 ({M_2} c_\beta+\mu  s_\beta)^2}{2
\left({M_2}^2-\mu ^2\right)^2}
\end{array}
\right) \left(1+\mathcal O \left(\frac{m_Z^2}{\mu^2}\right) \right)\ , \\
U_{ij}^{(e)} =& \ \delta_{ij}+\mathcal O \left( \zeta^2 \right) \ .
\end{align}
Numerically, the relative correction to the NLO contribution to $U_{\alpha\beta}^{(\chi^-)}$
is smaller than~$20\%$.
The off-diagonal elements of $U^{(c)}$ are
\begin{align}
U_{\alpha i}^{(\chi^-, e)} =& \ \zeta_i
\left(
\begin{array}{c}
 -\frac{ {m_Z}  c_w}{{M_2}} \\
 \frac{ {m_Z}^2  c_w^2 s_\beta}{{M_2} \mu }
\end{array}
\right)
\left(1 +\mathcal O \left(
\frac{s_{2 \beta} m_Z^2}{\mu^2}
\right)\right)\ , \\
U_{i \alpha }^{(e, \chi^- )} =& \ \zeta_i
\left(
\begin{array}{c}
 \frac{ {m_Z}  c_w}{{M_2}} \\
 \frac{ {m_Z}^2  c_w^2 (\mu  c_\beta+{M_2} s_\beta)}{\mu
^3-{M_2}^2 \mu }
\end{array}
\right)\left(1 +\mathcal O \left(
\frac{s_{2 \beta} m_Z^2}{\mu^2}
\right)\right)\ .
\end{align}
Here we ignored corrections that are proportional to the Yukawa
couplings $h_{ii}^e$ or higher powers thereof.
The numerical value of the higher order correction relative to the NLO term is
smaller than $1\%$ for $U_{\alpha i}^{(\chi^-, e)}$, smaller than $5\%$ for
$U_{i1 }^{(e, \chi^- )}$, and smaller than $15\%$ for
$U_{i2 }^{(e, \chi^- )}$.

\subsection{The currents in mass eigenstate basis}
The neutral and charged currents were given in Eqs.~(\ref{zmass})
and (\ref{wmass}),
\begin{align}
J_{Z \mu} =& \overline \chi^0_a \gamma_\mu V^{(\chi^0)}_{ab} \chi^0_b
+\overline \chi^-_\alpha  \gamma_\mu V^{(\chi^-)}_{\alpha\beta} \chi^-_\beta
+\overline \chi^+_\alpha \gamma_\mu V^{(\chi^+)}_{\alpha\beta} \chi^+_\beta
+\overline \nu_i \gamma_\mu V^{(\nu)}_{ij} \nu_i
+\overline e_i \gamma_\mu V^{(e)}_{ij} e_i	\nonumber \\
&+\left(\overline \chi^0_a \gamma_\mu V^{(\chi,\nu)}_{a i} \nu_i
+\overline \chi^-_\alpha \gamma_\mu V^{(\chi^-,e)}_{\alpha i} e_i
+\overline \chi^+_\alpha \gamma_\mu V^{(\chi^+,e^c)}_{\alpha i} {e_i^c}
+{\rm h.c.} \right) \ ,\nonumber \\
J_{ \mu}^- =& \
\overline \chi^0_a \gamma_\mu V^{(\chi)}_{a\alpha} \chi^-_\alpha
+\overline \chi^0_a \gamma_\mu V^{(\chi,e)}_{ai} e_i
+\overline \nu_i \gamma_\mu V^{(\nu,\chi)}_{i\alpha} \chi^-_\alpha
+\overline \nu_i \gamma_\mu V^{(\nu,e)}_{ij} e_j \ .\nonumber
\end{align}
The CKM-like matrices
$V^{(\chi^0)}_{ab}$,
$V^{(\chi^-)}_{\alpha\beta}$,
$V^{(\chi^+)}_{\alpha\beta}$,
$V^{(\nu)}_{ij}$,
$V^{(e)}_{ij}$,
$V^{(\chi,\nu)}_{a i}$,
$V^{(\chi^-,e)}_{\alpha i}$,
$V^{(\chi^+,e^c)}_{\alpha i}$,
$V^{(\chi)}_{a\alpha}$,
$V^{(\chi,e)}_{ai}$,
$V^{(\nu,\chi)}_{i\alpha}$,
$V^{(\nu,e)}_{ij}$
follow from the currents in gauge eigenbasis,
Eqs.~(\ref{zgauge}) and (\ref{wgauge}), and the explicit matrices
$U^{(n)}$, $U^{(c)}$ and $\widetilde U^{(c)}$. Here we focus on the matrices
$V^{(\chi,\nu)}_{a i}$ and $V^{(\chi,e)}_{ai}$ since they determine the
interactions of interest for this work.
We find
\begin{align}
V^{(\chi,\nu)}_{a i}=& \ \zeta_i
\left(
\begin{array}{c}
 -\frac{s_w {m_Z}  }{2 {M_1}} \\
 \frac{c_w {m_Z}  }{2 {M_2}} \\
 \frac{ {m_Z}^2 \mu v_1}{2 \sqrt{2} {M_1}
 ({M_1}-\mu ) ({M_2}-\mu )  } \\
 \frac{ {m_Z}^2 \mu v_2}{2 \sqrt{2} {M_1}
({M_1}+\mu ) ({M_2}+\mu )}
\end{array}
\right)
\left(1+\mathcal O\left( \begin{array}{c}
\frac{s_{2\beta}  m_Z^2}{\mu^2} \\
\frac{ m_Z^2}{ \mu^2}\\
\frac{ m_Z^2 }{\mu^2}\\
\frac{s_{2\beta}  m_Z^2}{ \mu^2}
\end{array}\right)\right) \ ,
\end{align}
with abbreviations
\begin{align}
v_1 =& \frac{1}{M_2 \mu^2} \left({M_1} ({M_1}-\mu ) ((\mu -2
{M_2}) c_\beta-\mu  s_\beta) c_w^2
\right. \nonumber \\ &\left.
+{M_2} ({M_2}-\mu ) ((\mu -2
{M_1}) c_\beta-\mu  s_\beta) s_w^2\right) \ , \\
v_2 =& \frac{1}{M_2 \mu^2}\left({M_1} ({M_1}+\mu ) ((2 {M_2}+\mu
) c_\beta+\mu  s_\beta) c_w^2
\right. \nonumber \\ &\left.
+{M_2} ({M_2}+\mu ) ((2 {M_1}+\mu )
c_\beta+\mu  s_\beta) s_w^2\right) \ .
\end{align}
Numerically, the relative errors are smaller than $0.10, 0.20, 0.15, 0.05$ for $a=1, \ldots, 4$.

Finally,
\begin{align}
V^{(\chi,e)}_{ai} =& \ \zeta_i
\left(
\begin{array}{c}
 -\frac{s_w {m_Z}  }{{M_1}} \\
 -\frac{\left(\sqrt{2}-1\right) c_w {m_Z}  }{{M_2}} \\
 -\frac{ m_Z^2 \mu \widetilde v_1}{ {M_2} ({M_1}-\mu )
({M_2}-\mu )  } \\
 \frac{m_Z^2 \mu \widetilde v_2}{ {M_2}
({M_1}+\mu ) ({M_2}+\mu )}
\end{array}
\right)
 \left(1+\mathcal O\left( \begin{array}{c}
 \frac{s_{2 \beta} m_Z^2}{\mu^2} \\
 \frac{ m_Z^2}{\mu^2} \\
 \frac{m_Z^2}{\mu^2}\\
\frac{m_Z^2}{\mu^2}
\end{array}\right) \right) \ ,
\end{align}
with abbreviations
\begin{align}
\widetilde v_1 =& \frac{1}{2 \mu^2}\left(c_\beta \left(({M_1}-\mu ) \left(\sqrt{2} {M_2}-2
\mu \right) c_w^2+\sqrt{2} {M_2} ({M_2}-\mu ) s_w^2\right)
\right.  \nonumber \\&\left.
+s_\beta \left(\left(\sqrt{2}-2\right) ({M_1}-\mu ) \mu  c_w^2+
\sqrt{2} {M_2} ({M_2}-\mu ) s_w^2\right)\right) \ , \\
\widetilde v_2 =& \frac{1}{2 \mu^2} \left(
c_\beta \left(({M_1}+\mu ) \left(\sqrt{2} {M_2}+2 \mu \right)
c_w^2+\sqrt{2} {M_2} ({M_2}+\mu ) s_w^2\right)
\right.  \nonumber \\&\left.
+s_\beta \left(\left(\sqrt{2}-2\right) \mu  ({M_1}+\mu ) c_w^2-\sqrt{2} {M_2} ({M_2}+\mu ) s_w^2\right)\right) \ .
\end{align}
Here we again neglected corrections that involve the Yukawa couplings $h_{ii}^e$.
The numerical corrections to the NLO contributions to $V^{(\chi,e)}_{ai}$
 are smaller than
$0.05, 0.15, 0.20$ for $a=1, 2, 3$, respectively.
For $a=4$ we reach the limit of our numerical precision.

\end{appendix}


\newpage

\begin{thebibliography}{99}

\bibitem{fnf76}
  D.~Z.~Freedman, P.~van Nieuwenhuizen and S.~Ferrara,
  Phys.\ Rev.\  D {\bf 13} (1976) 3214;\\
  S.~Deser and B.~Zumino,
  Phys.\ Lett.\  B {\bf 62} (1976) 335.

\bibitem{pp81}
  H.~Pagels and J.~R.~Primack,
  Phys.\ Rev.\ Lett.\  {\bf 48} (1982) 223.

\bibitem{we82}
  S.~Weinberg,
  Phys.\ Rev.\ Lett.\  {\bf 48} (1982) 1303.

\bibitem{ens85}
  J.~R.~Ellis, D.~V.~Nanopoulos and S.~Sarkar,
  Nucl.\ Phys.\  B {\bf 259} (1985) 175.

\bibitem{kkm05}
M.~Kawasaki, K.~Kohri and T.~Moroi,
Phys.\ Lett.\ B {\bf 625}, 7 (2005)
[astro-ph/0402490];
Phys.\ Rev.\ D {\bf 71}, 083502 (2005)
[astro-ph/0408426];\\
  K.~Jedamzik,
  Phys.\ Rev.\  D {\bf 74}, 103509 (2006)
  [hep-ph/0604251].

\bibitem{fy86}
  M.~Fukugita and T.~Yanagida,
  Phys.\ Lett.\  B {\bf 174}, 45 (1986).

\bibitem{bbp98}
  M.~Bolz, W.~Buchm\"uller and M.~Plumacher,
  Phys.\ Lett.\  B {\bf 443} (1998) 209
  [hep-ph/9809381].

\bibitem{fen05}
For a recent review and references, see\\
J.~L.~Feng,
{\it Supersymmetry and cosmology},
Annals Phys.\  {\bf 315} (2005) 2.

\bibitem{bcx07}
  W.~Buchm\"uller, L.~Covi, K.~Hamaguchi, A.~Ibarra and T.~Yanagida,
  JHEP {\bf 0703} (2007) 037
  [hep-ph/0702184].

\bibitem{ty00}
  F.~Takayama and M.~Yamaguchi,
  Phys.\ Lett.\  B {\bf 485}, 388 (2000)
  [hep-ph/0005214].

\bibitem{lor07}
  S.~Lola, P.~Osland and A.~R.~Raklev,
  Phys.\ Lett.\  B {\bf 656} (2007) 83
  [0707.2510 [hep-ph]].

\bibitem{bbx07}
  G.~Bertone, W.~Buchm\"uller, L.~Covi and A.~Ibarra,
  JCAP {\bf 0711} (2007) 003
  [0709.2299 [astro-ph]].

\bibitem{it07}
  A.~Ibarra and D.~Tran,
  Phys.\ Rev.\ Lett.\  {\bf 100}, 061301 (2008)
  [0709.4593 [astro-ph]].

\bibitem{imm08}
  K.~Ishiwata, S.~Matsumoto and T.~Moroi,
  Phys.\ Rev.\  D {\bf 78} (2008) 063505
  [0805.1133 [hep-ph]].

\bibitem{bix09}
  W.~Buchm\"uller, A.~Ibarra, T.~Shindou, F.~Takayama and D.~Tran,
  JCAP {\bf 0909} (2009) 021
  [0906.1187 [hep-ph]].

\bibitem{blx09}
  N.~E.~Bomark, S.~Lola, P.~Osland and A.~R.~Raklev,
  Phys.\ Lett.\  B {\bf 686} (2010) 152
  [0911.3376 [hep-ph]].

\bibitem{crx10}
  K.~Y.~Choi, D.~Restrepo, C.~E.~Yaguna and O.~Zapata,
  [1007.1728 [hep-ph]].

\bibitem{FermiLAT1}
  A.~A.~Abdo {\it et al.},
  Phys.\ Rev.\ Lett.\  {\bf 104} (2010) 091302
  [1001.4836 [astro-ph.HE]].

\bibitem{FermiLAT2}
  A.~A.~Abdo {\it et al.}  [The Fermi-LAT collaboration],
  Phys.\ Rev.\ Lett.\  {\bf 104} (2010) 101101
  [1002.3603 [astro-ph.HE]].

\bibitem{bes08}
  W.~Buchm\"uller, M.~Endo, T.~Shindou,
  JHEP {\bf 0811} (2008) 079
  [0809.4667 [hep-ph]].

\bibitem{hs84}
  L.~J.~Hall and M.~Suzuki,
  Nucl.\ Phys.\ B {\bf 231} (1984) 419.

\bibitem{add04}
For recent discussions and references, see\\
  B.~C.~Allanach, A.~Dedes and H.~K.~Dreiner,
  Phys.\ Rev.\ D {\bf 69} (2004) 115002
  [Erratum-ibid.\ D {\bf 72} (2005) 079902],
  [hep-ph/0309196];\\
  R.~Barbier {\it et al.},
  Phys.\ Rept.\  {\bf 420} (2005) 1
  [hep-ph/0406039];\\
  F.~de~Campos \textit{et al.},
  JHEP \textbf{05} (2008) 048
  [hep-ph/07122156].

\bibitem{softsusy}
  B.~C.~Allanach et al.,
  Comput.\ Phys.\ Commun.\  {\bf 181} (2010) 232
  [0903.1805 [hep-ph]].

\bibitem{cdx91}
  B.~A.~Campbell, S.~Davidson, J.~R.~Ellis, K.~A.~Olive,
  Phys.\ Lett.\ B {\bf 256} (1991) 484;\\
  W.~Fischler, G.~F.~Giudice, R.~G.~Leigh and S.~Paban,
  Phys.\ Lett.\ B {\bf 258} (1991) 45;\\
  H.~K.~Dreiner and G.~G.~Ross,
  Nucl.\ Phys.\ B {\bf 410} (1993) 188
  [hep-ph/9207221].

\bibitem{ehi09}
  M.~Endo, K.~Hamaguchi and S.~Iwamoto,
  JCAP {\bf 1002} (2010) 032
  [0912.0585 [hep-ph]].

\bibitem{gm88}
  G.~F.~Giudice and A.~Masiero,
  Phys.\ Lett.\ B {\bf 206} (1988) 480.

\bibitem{by99}
  W.~Buchm\"uller and T.~Yanagida,
  Phys.\ Lett.\ B {\bf 445} (1999) 399
  [hep-ph/9810308].

\bibitem{bdh99}
  W.~Buchm\"uller, D.~Delepine and L.~T.~Handoko,
  Nucl.\ Phys.\  B {\bf 576} (2000) 445
  [hep-ph/9912317].

\bibitem{gr08}
M.~Grefe, DESY-THESIS-2008-043

\bibitem{mrv98}
  B.~Mukhopadhyaya, S.~Roy and F.~Vissani,
  Phys.\ Lett.\ B {\bf 443} (1998) 191
  [hep-ph/9808265];\\
  E.~J.~Chun and J.~S.~Lee,
  Phys.\ Rev.\ D {\bf 60}, 075006 (1999)
  [hep-ph/9811201].

\bibitem{iim08}
  K.~Ishiwata, T.~Ito and T.~Moroi,
  Phys.\ Lett.\  B {\bf 669} (2008) 28
  [0807.0975 [hep-ph]].

\bibitem{ahs09}
  S.~Asai, K.~Hamaguchi and S.~Shirai,
  Phys.\ Rev.\ Lett.\  {\bf 103} (2009) 141803
  [0902.3754 [hep-ph]].
%
\bibitem{bhx04}
  W.~Buchm\"uller, K.~Hamaguchi, M.~Ratz, T.~Yanagida,
  \mbox{Phys.\ Lett.\ B {\bf 588} (2004) 90}
  [hep-ph/0402179].

\end{thebibliography}
\end{document}